\documentclass{article}
\usepackage[utf8]{inputenc}
\usepackage{amsmath, amsfonts}
\usepackage{xcolor}
\usepackage{hyperref}
\usepackage{graphicx}
\usepackage{geometry}
\usepackage{placeins}
\usepackage{authblk}
\usepackage[pagewise]{lineno}

\hypersetup{
        bookmarksnumbered,
        unicode,                        
        colorlinks,                     
        citecolor=[rgb]{.9,0,.5},       
        urlcolor=[rgb]{0,0,1},  
        linkcolor=[rgb]{0,.7,0} 
        }

\usepackage[
    natbib=true, style=numeric,
    sorting=none
]{biblatex}

\usepackage{tikz}
 
\usetikzlibrary{shapes.misc}

 \tikzset{cross/.style={cross out, draw=black, minimum size=2*(#1-\pgflinewidth), inner sep=0pt, outer sep=0pt},
cross/.default={2pt}}

\usepackage{import}

\newcommand{\bref}[1]{(\ref{#1})}
\newcommand{\pd}{\partial}

\newmuskip\pFqmuskip

\newcommand*\pFq[6][8]{%
  \begingroup 
  \pFqmuskip=#1mu\relax
  \mathcode`\,=\string"8000
  \begingroup\lccode`\~=`\,
  \lowercase{\endgroup\let~}\pFqcomma
  {}_{#2}F_{#3}{\left[\genfrac..{0pt}{}{#4}{#5};#6\right]}%
  \endgroup
}
\newcommand{\pFqcomma}{\mskip\pFqmuskip}

\newcommand*\pGq[3][8]{%
  \begingroup 
  \pFqmuskip=#1mu\relax
  \mathcode`\,=\string"8000
  \begingroup\lccode`\~=`\,
  \lowercase{\endgroup\let~}\pFqcomma
  \Gamma{\left[\genfrac..{0pt}{}{#2}{#3}\right]}%
  \endgroup
}

\title{Triple Product Amplitude from Chiral String}

\author[$1, 2$]{ Yu-Ping Wang \footnote{Email: \href{yu-ping.wang@stonybrook.edu}{yu-ping.wang@stonybrook.edu} } }

\date{\today}
\affil[$1$]{\textit{ Department of Physics, SUNY Stony Brook University, Stony Brook, NY 11794, USA}}
\affil[$2$]{\textit{C. N. Yang Institute for Theoretical Physics. Stony Brook, NY 11794, USA}}

\begin{document}

\hfill YITP-SB-2024-17
{\let\newpage\relax\maketitle}

\begin{abstract}
In this paper, we proposed a worldsheet construction of a subset of triple product amplitudes proposed by Huang and Remmen (2022). We start with closed bosonic strings but left and right-moving momenta are not necessarily equal. Instead, they satisfy certain conditions. We called them section conditions. These conditions are generalizations of the section condition in double field theory.  The vertex operators of chiral strings have nontrivial monodromy, so we interpret them as attached to the end of defects. In the calculation of the amplitude, we not only have to integrate over the moduli space, we also have to sum over different defect configurations. 

Unitarity and other consistency conditions for chiral string amplitudes are checked. We found the graviton amplitude, the Virasoro amplitude, and also a special kind of amplitude that has one infinite spin tower. Similar kinds of amplitude have appeared in bootstrap literature. The more general $N$-point amplitude could be obtained from a modified KLT relation. The five-point chiral string amplitude is also explicitly calculated.

\end{abstract}

\newpage
\tableofcontents
\section{Introduction}

 In recent years, there has been great interest in understanding Yang-Mills and graviton amplitude using techniques developed from string theories. The famous KLT relations \cite{Kawai:1985xq} and BCJ relations \cite{Bern:2008qj} could both be derived from world-sheet integrals \cite{Stieberger:2009hq, Bjerrum-Bohr:2009ulz}. 
 
Inspired by the relations above, one might wonder if it is possible to construct some sort of world-sheet integral representations of Yang-Mills or gravitons amplitudes. Much progress has been made in this direction. CHY formalism, pioneered by Cachazo, He, and Yuan, is a prescription for massless particle amplitudes in arbitrary dimensions, where the amplitude is represented as an integration over moduli space of n-punctured sphere \cite{Cachazo:2013gna, Cachazo:2013hca, Cachazo:2013iea}. Later, the ambitwistor string was developed by Mason and Skinner. In which they take the chiral infinite tension limit of usual strings, and find that only the massless spectrum survives, and its quantization also leads to the same scatter equation from CHY formalism \cite{Mason:2013sva}. 

Later, an alternative formalism was developed by Siegel, Huang and Ellis, which they call ``chiral string''. It directly calculates graviton amplitudes by a twisted version of the KLT formula \cite{Siegel:2015axg, Huang:2016bdd}. 

In chiral string formalism, instead of taking the chiral infinite tension limit, one could simply change the boundary condition for the string propagator. 
\begin{equation}\label{dprop}
G(z, \bar{z}) = \frac{1}{2}\ln(z\bar{z}) \rightarrow \frac{1}{2}\left(\ln(z) - \ln(\bar{z})\right). 
\end{equation}
Plugging this propagator into the standard derivation of the KLT formula, we would get a modified version of the KLT formula

\begin{equation}
    A_{\textrm{open}}^{T} \cdot S  \cdot A_{\textrm{open}} \quad \Rightarrow  \quad A_{\textrm{open}}^{T}  \cdot S  \cdot \tilde{A}_{\textrm{open}} \equiv M_{\textrm{chrial}}.
\end{equation}
$M_{\textrm{chiral}}$ is the $n$-point usual chiral string amplitude computed with the modified propagator. $A_{\textrm{open}}$ are basis of  $(n-3)!$ independent open string $n$-point amplitudes, while  $\tilde{A}_{\textrm{open}}$ are same amplitudes but with the sign of string tension $\alpha{}'$ flipped. $S$ is the momentum kernel in KLT relation written as a $(n-3)!\times (n-3)!$ matrix \cite{Bjerrum-Bohr:2010pnr}.

It is proved in \cite{Huang:2016bdd} that chiral string amplitudes compute graviton amplitudes.\footnote{ They proved this statement perturbatively in inverse string tension $\alpha{}'$, although it relies on some conjecture on properties of multi-zeta values (MZVs) expansions of open string amplitudes \cite{Broedel:2013tta, Schlotterer:2012ny}.} 

In this paper, we aim to build upon chiral string formalism. First, we found that modification of propagator \bref{dprop} is equivalent to changing vertex operators such that left and right moving momentum are independent, and then imposed a modified version of section condition.
More specifically, we considered vertex operators with exponential factors.
$$
 e^{i k_{L}\cdot X_{L}(z) + i k_{R}\cdot X_{R}(\bar{z}) }.
$$

There are two problems with this approach. 
First dimension of momenta have doubled as we put $K^{M} = (\,k_{R}^{\mu}\,, \,k_{L}^{\mu}\,)$. To reduce the dimension back to its original value, we need to impose some sort of constraint on the doubled momentum.

Second, since left and right-moving momenta are now independent of each other,  correlators of these vertex operators will produce cuts on the worldsheet. We interpret these cuts as defects on the worldsheet, and they might cause obstructions in making amplitudes crossing symmetric. 

\begin{figure}[h]
    \centering
        \caption{ Defects on the worldsheet will cause a phase factor when one vertex operator goes around another.  \label{pics6}}
    \tikzset{every picture/.style={line width=0.5pt}} 

\begin{tikzpicture}[x=0.5pt,y=0.5pt,yscale=-1,xscale=1]

\draw    (371.5,130.8) -- (380.3,140.33) ;
\draw    (380.7,130.93) -- (371.5,140.33) ;

\draw  [fill={rgb, 255:red, 0; green, 0; blue, 0 }  ,fill opacity=1 ] (159.98,69.5) .. controls (159.98,68.67) and (160.65,68) .. (161.48,68) .. controls (162.31,68) and (162.98,68.67) .. (162.98,69.5) .. controls (162.98,70.33) and (162.31,71) .. (161.48,71) .. controls (160.65,71) and (159.98,70.33) .. (159.98,69.5) -- cycle ;
\draw  [dash pattern={on 4.5pt off 4.5pt}]  (161.48,71) -- (161.31,105.14) -- (160.75,218.47) ;
\draw [shift={(160.74,220.47)}, rotate = 270.28] [color={rgb, 255:red, 0; green, 0; blue, 0 }  ][line width=0.75]    (10.93,-3.29) .. controls (6.95,-1.4) and (3.31,-0.3) .. (0,0) .. controls (3.31,0.3) and (6.95,1.4) .. (10.93,3.29)   ;
\draw    (36.74,134.47) .. controls (38.41,132.8) and (40.08,132.81) .. (41.74,134.48) .. controls (43.4,136.15) and (45.07,136.16) .. (46.74,134.5) .. controls (48.41,132.84) and (50.08,132.85) .. (51.74,134.52) .. controls (53.4,136.19) and (55.07,136.2) .. (56.74,134.54) .. controls (58.41,132.88) and (60.08,132.89) .. (61.74,134.56) .. controls (63.41,136.23) and (65.07,136.23) .. (66.74,134.57) .. controls (68.41,132.91) and (70.08,132.92) .. (71.74,134.59) .. controls (73.4,136.26) and (75.07,136.27) .. (76.74,134.61) .. controls (78.41,132.95) and (80.08,132.96) .. (81.74,134.63) .. controls (83.41,136.3) and (85.07,136.3) .. (86.74,134.64) .. controls (88.41,132.98) and (90.08,132.99) .. (91.74,134.66) .. controls (93.4,136.33) and (95.07,136.34) .. (96.74,134.68) .. controls (98.41,133.02) and (100.08,133.03) .. (101.74,134.7) .. controls (103.41,136.37) and (105.07,136.37) .. (106.74,134.71) .. controls (108.41,133.05) and (110.08,133.06) .. (111.74,134.73) .. controls (113.4,136.4) and (115.07,136.41) .. (116.74,134.75) .. controls (118.41,133.09) and (120.08,133.1) .. (121.74,134.77) .. controls (123.4,136.44) and (125.07,136.45) .. (126.74,134.79) .. controls (128.41,133.13) and (130.07,133.13) .. (131.74,134.8) .. controls (133.4,136.47) and (135.07,136.48) .. (136.74,134.82) .. controls (138.41,133.16) and (140.08,133.17) .. (141.74,134.84) .. controls (143.4,136.51) and (145.07,136.52) .. (146.74,134.86) .. controls (148.41,133.2) and (150.07,133.2) .. (151.74,134.87) .. controls (153.4,136.54) and (155.07,136.55) .. (156.74,134.89) .. controls (158.41,133.23) and (160.08,133.24) .. (161.74,134.91) .. controls (163.4,136.58) and (165.07,136.59) .. (166.74,134.93) .. controls (168.41,133.27) and (170.07,133.27) .. (171.74,134.94) .. controls (173.4,136.61) and (175.07,136.62) .. (176.74,134.96) .. controls (178.41,133.3) and (180.08,133.31) .. (181.74,134.98) .. controls (183.4,136.65) and (185.07,136.66) .. (186.74,135) .. controls (188.41,133.34) and (190.08,133.35) .. (191.74,135.02) .. controls (193.41,136.69) and (195.07,136.69) .. (196.74,135.03) .. controls (198.41,133.37) and (200.08,133.38) .. (201.74,135.05) .. controls (203.4,136.72) and (205.07,136.73) .. (206.74,135.07) .. controls (208.41,133.41) and (210.08,133.42) .. (211.74,135.09) .. controls (213.41,136.76) and (215.07,136.76) .. (216.74,135.1) .. controls (218.41,133.44) and (220.08,133.45) .. (221.74,135.12) .. controls (223.4,136.79) and (225.07,136.8) .. (226.74,135.14) .. controls (228.41,133.48) and (230.08,133.49) .. (231.74,135.16) .. controls (233.41,136.83) and (235.07,136.83) .. (236.74,135.17) .. controls (238.41,133.51) and (240.08,133.52) .. (241.74,135.19) .. controls (243.4,136.86) and (245.07,136.87) .. (246.74,135.21) .. controls (248.41,133.55) and (250.08,133.56) .. (251.74,135.23) .. controls (253.4,136.9) and (255.07,136.91) .. (256.74,135.25) .. controls (258.41,133.59) and (260.07,133.59) .. (261.74,135.26) .. controls (263.4,136.93) and (265.07,136.94) .. (266.74,135.28) .. controls (268.41,133.62) and (270.08,133.63) .. (271.74,135.3) .. controls (273.4,136.97) and (275.07,136.98) .. (276.74,135.32) .. controls (278.41,133.66) and (280.07,133.66) .. (281.74,135.33) .. controls (283.4,137) and (285.07,137.01) .. (286.74,135.35) .. controls (288.41,133.69) and (290.08,133.7) .. (291.74,135.37) .. controls (293.4,137.04) and (295.07,137.05) .. (296.74,135.39) .. controls (298.41,133.73) and (300.07,133.73) .. (301.74,135.4) .. controls (303.4,137.07) and (305.07,137.08) .. (306.74,135.42) .. controls (308.41,133.76) and (310.08,133.77) .. (311.74,135.44) .. controls (313.4,137.11) and (315.07,137.12) .. (316.74,135.46) .. controls (318.41,133.8) and (320.08,133.81) .. (321.74,135.48) .. controls (323.41,137.15) and (325.07,137.15) .. (326.74,135.49) .. controls (328.41,133.83) and (330.08,133.84) .. (331.74,135.51) .. controls (333.4,137.18) and (335.07,137.19) .. (336.74,135.53) .. controls (338.41,133.87) and (340.08,133.88) .. (341.74,135.55) .. controls (343.41,137.22) and (345.07,137.22) .. (346.74,135.56) .. controls (348.41,133.9) and (350.08,133.91) .. (351.74,135.58) .. controls (353.4,137.25) and (355.07,137.26) .. (356.74,135.6) .. controls (358.41,133.94) and (360.08,133.95) .. (361.74,135.62) .. controls (363.4,137.29) and (365.07,137.3) .. (366.74,135.64) .. controls (368.41,133.98) and (370.07,133.98) .. (371.74,135.65) -- (375.64,135.67) -- (375.64,135.67) ;
\draw  [fill={rgb, 255:red, 0; green, 0; blue, 0 }  ,fill opacity=1 ] (159.24,220.47) .. controls (159.24,219.64) and (159.92,218.97) .. (160.74,218.97) .. controls (161.57,218.97) and (162.24,219.64) .. (162.24,220.47) .. controls (162.24,221.3) and (161.57,221.97) .. (160.74,221.97) .. controls (159.92,221.97) and (159.24,221.3) .. (159.24,220.47) -- cycle ;

\draw (387.64,124.67) node [anchor=north west][inner sep=0.75pt]   [align=left] {$\displaystyle V_{1}$};
\draw (166.64,54.33) node [anchor=north west][inner sep=0.75pt]   [align=left] {$\displaystyle V_{2}$};
\draw (169.98,205.67) node [anchor=north west][inner sep=0.75pt]   [align=left] {$\displaystyle e^{i\pi K_{1} \cdot K_{2}} V_{2}$};

\end{tikzpicture}
\end{figure}
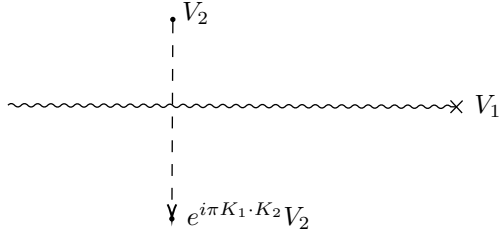

Both problems could be solved if we impose a section condition on momenta. Given two vertex operators with doubled momenta. Suppose we have two vertex operators that have doubled momentum $k_{R}, k_{L}$ and $k_{R}{}', k_{L}{}'$. If 
 the first vertex operators go around the second one, it will gain an extra phase $\pi(k_{R}{}'\cdot k_{R}-k_{L}{}'\cdot k_{L}) \equiv \pi K\cdot K{}'$. See figure \ref{pics6}.

 The inner product between $K$ and $K{}'$ is just the standard $O(d, d)$ inner product. If we don't want cuts in correlators, then we can set $K\cdot K{}' = 0$ for every pair of doubled momenta in amplitudes. This is nothing but the famous section condition for double field theory (equivalently, the two vierbein formalism for strings)\cite{Siegel:1993xq, Siegel:1993th, Hull:2009mi, Aldazabal:2013sca}.
Solving this section condition, we would reduce the dimensions of momenta back to $d$.

By applying this section condition, we arrived at the usual string amplitude, but if we used the following section condition instead.
$$k_{R}{}'\cdot k_{R}+k_{L}{}'\cdot k_{L}=0$$
We arrived at the chiral strings proposed by Siegel, Huang and Ellis
$$
\frac{1}{2}k_{R}{}'\cdot k_{R}\ln z + \frac{1}{2}k_{L}{}'\cdot k_{L}\ln \bar{z} \quad \Rightarrow\quad 
\frac{1}{2}k\cdot k{}'(\ln z -\ln \bar{z}) .
$$

Note that there is a further generalization for section conditions. If we want to avoid branch cuts in correlators, the usual section condition is too restrictive. We just need to impose 
\begin{equation}\label{sectioncond}
k_{R} \cdot k_{R}{}' - k_{L} \cdot k_{L}{}'= 2n, \quad \textrm{$n$ is an integer.}
\end{equation}
In the rest of the paper, we shall call it the standard section condition.

In this paper, we also proposed another section condition 
\begin{equation}\label{sectioncond2}
k_{R} \cdot k_{R}{}' + k_{L} \cdot k_{L}{}'= 2n, \quad \textrm{$n$ is an integer.}
\end{equation}

We call this the twisted section condition.

By design, standard section conditions will result in crossing symmetric amplitudes. Surprisingly, we find that amplitudes constructed using twisted section conditions are also crossing symmetric. 
For $n = 0$, standard and twisted section conditions lead to string and graviton amplitudes, and in this paper, we would like to initiate the investigation when more general section conditions described above are imposed.   

We shall contend ourselves with 4-point amplitudes constructed from vertex operators of the form $V = \partial X \bar{\partial} X e^{iK\cdot X}$, and demanded that it have a massless spin-2 exchange (graviton) and no tachyons. We found that these amplitudes are all special cases of triple-product amplitudes proposed in \cite{Huang:2022mdb}.

In \cite{Huang:2022mdb}, Huang and Remmen observed that amplitudes with the following ansatz 
$$
M(s, t, u) = A(s)A(t)A(u), 
$$
where
$$
A(s) = \frac{1}{s} + \sum_{a= 1}\frac{g_{a}^{2}}{-s + m_{a}^2}, \quad $$
satisfy both unitarity constraint and UV-completeness if $\sum g_{a}^{2} = 1.
$

In particular, we found that 4-point chiral string amplitude may be characterized by a positive integer $k$.
$$
M_{\textrm{Chiral}}(s, t, u) = \mathcal{W}^{4}F_{k}(s)F_{k}(t)F_{k}(u),
$$ 
where
$$
F_{k}(x)= \frac{1}{(1+s)_{k}}\frac{\Gamma(-s)}{\Gamma(1+s)},
$$
for standard section conditions. Here $(s)_{x}\equiv s(s + 1)\cdots (s +x -1)$ is the Pochhammer Symbol. 

$$
F_{k}(s) = \frac{1}{(s - k)_{k + 1}} = \frac{1}{s(s-1)\cdots(s-k)},
$$
for twisted section condition case. $\mathcal{W}^{4}$ is the universal kinematic prefactors that consist of contractions of four Weyl tensors.\footnote{
$
\mathcal{W}^4 =32 (W_1)^{\mu\nu\rho\sigma}(W_2)_{\mu\;\;\rho}^{\;\;\alpha\;\;\beta}(W_3)_{\alpha\;\;\nu}^{\;\;\gamma\;\;\delta}(W_4)_{\beta\gamma\sigma\delta} - 8 (W_1)^{\mu\nu\rho\sigma}(W_2)_{\mu\nu}^{\;\;\;\;\alpha\beta}(W_3)_{\rho\alpha}^{\;\;\;\;\gamma\delta}(W_4)_{\sigma\beta\gamma\delta}
$
plus symmetric permutation of labels $1, 2, 3, 4$. $(W_{i})_{\mu\nu}{}^{\rho\sigma} = p_{i\, [\mu}\,p_{i}^{[\rho}\,(\epsilon_{i})_{\nu]}{}^{\sigma]}$ is linearized Weyl tensor for $i$th particles. Note that when momenta are on-shell, the linearized Weyl tensor is identical to the linearized Riemann tensor.}

As noted in the original paper \cite{Huang:2022mdb}, residues in amplitudes are generally not polynomials in scattering angle $x = \cos\theta$. That means resonance state exchange for a particular mass contains infinite spins. One interpretation of an infinite spin tower is amplitudes are exchanging of some extended object.

Unfortunately, our chiral string amplitude did not satisfy $\sum g_{a}^{2} = 1$, so the unitarity condition needs to be separately checked. We find out that only new amplitudes that satisfy unitarity constraints are amplitudes with twisted section condition and $k = 1$.
\begin{equation}\label{sptwamp}
\frac{\mathcal{W}^{4}}{s(s-1)t(t-1)u(u-1)}.
\end{equation} 

This amplitude has only one fixed mass but has infinite numbers of spin exchanges on that mass. Amplitudes with this property have been found in the context of EFT bootstrap \cite{Berman:2023jys, Albert:2022oes,Cheung:2024obl}. They call such an amplitudes infinite spin tower (IST). We shall refer to this amplitude above as Virasoro IST.

Infinite spin towers are usually found in the cusp of allowed regions of theory space, but it is usually considered an unphysical theory. The fact that there is a worldsheet description of it might suggest that those ISTs are not necessarily unphysical. 

We should also note that we do not exhaust all possibilities of a unitary amplitude from chiral strings. Since we are only using vertex operators of type $V = \partial X \bar{\partial} X e^{iK\cdot X}$. To have a more exhaustive search one needs to have a BRST prescription of some sort and find all vertex operators that sit in BRST cohomology class. This will be left for future work. 
\subsection{Summery}

In section \ref{sec:general-formalism}, we establish the general formalism of calculating chiral string amplitude. When left-moving and right-moving momenta are different, there would in general be non-trivial monodromy around vertex operators.  We interpreted this as defects attaching at the end of vertex operators.  See figure \ref{pics1}. Therefore, to get full amplitude, one must sum over all possible topological inequivalent configurations of defects.  
By analytically continuing left-moving and right-moving coordinates and being careful about $i\epsilon$ prescription of string propagators, one arrived at a modified version of KLT relation \bref{mklt} that gives chiral string amplitude.
In section \ref{sec:Four-point-amplitudes} we focus on constructing  4-point chiral string amplitudes. We need to sum over 3 independent configurations to get a manifestly crossing invariant amplitude. 

In section \ref{sec:section-conditions}, we found that when certain conditions on left and right-moving momenta are met. The three different configurations in the previous section become identical, and the 4-point amplitude will collapse into a single term. There are two types of section conditions. The standard section condition \bref{sectioncond} and the twisted section condition \bref{sectioncond2}. The Virasoro amplitude is a special case of standard section condition while the 4-graviton amplitude is a special case of twisted section condition.
In section \ref{sec:soulutions-for-section-conditions}, we systematically find the solutions of the standard and twisted section conditions. We distinguish between the vertex operator momentum $K^{M}$ and the physical momentum $p^{\mu}$. The most general solution of $K^{M}$ \bref{secsol} contains a continuous part and a discrete part, which we interpret as the physical momenta and the winding modes.

In section \ref{sec:triple-product-amplitudes}, we study the 4-point amplitude under section conditions in more detail. We found that it is always in triple product form $M(s, t, u) = {\cal W}^4F(s)F(t)F(u)$. For amplitudes under standard section conditions, there is an infinite number of resonance states, but only a finite of them are infinite spin towers. For amplitude under the twisted section condition, there are three distinct cases. 1. The amplitude is a pure polynomial with no poles. It represents theories with only higher derivative contact terms. 2. Amplitudes with a finite number of resonance poles, but each of them is an infinite spin tower, this is the most interesting case. 3. The marginal case where it is graviton amplitude. 

In section \ref{sec:unitarity-constraints}, we study the consistency of the triple product amplitudes given in the previous section. The necessary conditions for amplitudes to be consistent are the unitarity constraints, the UV-completeness (Softness at fix-angle large scattering), and satisfy Regge bound. 
The only new type of amplitude that we find satisfies all these conditions is given at \bref{sptwamp}. This type of amplitude seems to occur naturally in amplitude bootstrap. In section \ref{sec:more-general-amplitudes}, we try to deform the triple product amplitude we found in the previous section. If we multiply them with certain polynomials of Mandelstam variables, we will get a large class of consistent amplitudes, which all exhibit infinite spin towers. We argue that these amplitudes could also be obtained from chiral string worldsheets using different vertex operators.

In section \ref{sec:higher-point-amplitude} we study in more detail general $N$-point amplitude. Under section conditions, the modified KLT relation simplifies to a bilinear product of $(N-3)!\times (N-3)!$ left and right moving open string amplitudes. See \bref{modNKLT}. In section \ref{sec:5-point-amplitudes} we give an explicit result of 5-point chiral string amplitudes in terms of ${}_3F_2$ using the simplified version of the KLT relation given previously.

\section{General formalism}\label{sec:general-formalism}

In this section, we shall start by extending the usual vertex operators in string theories to make left-moving and right-moving momentum independent and get the modified KLT relations. For simplicity, we shall only consider vertex operators for bosonic strings, but generalization to superstrings is trivial.

In bosonic string theory, we can decompose free boson field into left and right moving parts.
$$
X^{\mu}(z, \bar{z}) = X^{\mu}_{L}(z) + X^{\mu}_{R}(\bar{z}).
$$

Where OPEs of $X_{L}, X_{R}$ are
\begin{equation}
X_{R}^{\mu}(z)X_{R}^{\nu}(w) \sim -\frac{\alpha^{'}}{2}\eta^{\mu\nu}\ln (z - w), \quad 
X_{L}^{\mu}(\bar{z})X_{L}^{\nu}(\bar{w}) \sim -\frac{\alpha^{'}}{2}\eta^{\mu\nu}\ln (\bar{z} - \bar{w})
\end{equation}

 Asymptotic states of string theories are described by their vertex operators.  The most general forms of vertex operators are
$$
V^{(s, s^{'})}(z; k) = \epsilon^{(s)}_{\mu_{1}\cdots \mu_{2}}{}^{\nu_{1}\cdots \nu_{m}} \partial X_{L}^{\mu_{1}} \cdots X_{L}^{\mu_{n}}\bar{\partial} X_{R\;\nu_{1}} \cdots \bar{\partial} X_{R\;\nu_{n}} e^{ik\cdot (X_{R}(z) + X_{L}(\bar{z}))}
$$

 Hilbert space of string theory is extended by including vertex operators that have independent left and right moving momenta.

\begin{equation}\label{vertexop}
e^{ik \cdot X(z, \bar{z})} \mapsto e^{i k_{L}\cdot X_{L}(z) + i k_{R}\cdot X_{R}(\bar{z}) }
\end{equation}

$d$-dimensional left and right moving momentum are organized into a $2d$ momentum $K^{M} \equiv (\,k_{R}^{\mu}, \; k_{L}^{\nu}\,)$. 
For brevity, the exponential factor in vertex operators is written as $e^{iX_{M}\cdot X^{M}}$, which has manifest $T$-duality symmetry $O(d, d)$\cite{Duff:1989tf}.  

Any two of the extended vertex operators $V(z; \, k_{1L}, k_{1R})$, $V(z; \, k_{2L}, k_{2R})$ have non-trivial relative monodromy. From their OPE
$$
V(z_{1}; \, k_{1L}, k_{1R}) V(z_{2}; \, k_{2L}, k_{2R}) \sim (\bar{z}_{12})^{-\frac{\alpha{}'}{2}k_{1L}\cdot k_{2L}}(z_{12})^{-\frac{\alpha{}'}{2}k_{1R}\cdot k_{2R}}.
$$
Thus, phase factor from monodromy is $\frac{\alpha{}'}{2}(k_{1L}\cdot k_{2L} - k_{1R}\cdot k_{2R})$. If we want monodromy to vanish between vertex operators, the following constraints on momenta need to be imposed.

$$
k_{1L}\cdot k_{2L} - k_{1R}\cdot k_{2R}  \equiv K_{1}^{M}\eta_{MN}K_{2}^{N} = \frac{4n}{\alpha{}'}, \quad n \in \mathbb{Z},
$$
where $\eta_{MN} = \begin{pmatrix} I & 0 \\ 0 & -I \\ \end{pmatrix} $ is the $O(d, d)$ invariant matrix. This constraint is a slight generalization of section conditions for manifestly $T$-dual theory\cite{Siegel:1993bj, Siegel:1993th, Tseytlin:1990nb, Tseytlin:1990va}.

One might think of section condition \bref{sectioncond} as a consistency condition that needs to be imposed on our world-sheet theory. But as we shall see, even in some cases when \bref{sectioncond} is violated, we still get some physically interesting amplitude. In particular, if we impose instead 
$$    
k_{1L}\cdot k_{2L} + k_{1R}\cdot k_{2R} = 0 .
$$
We would get \emph{graviton} amplitudes instead of string amplitude.

\begin{figure}[h]
    \centering
    \tikzset{every picture/.style={line width=0.75pt}} 

\begin{tikzpicture}[x=0.75pt,y=0.75pt,yscale=-1,xscale=1]

\draw   (50,150) .. controls (50,94.77) and (94.77,50) .. (150,50) .. controls (205.23,50) and (250,94.77) .. (250,150) .. controls (250,205.23) and (205.23,250) .. (150,250) .. controls (94.77,250) and (50,205.23) .. (50,150) -- cycle ;
\draw  [fill={rgb, 255:red, 0; green, 0; blue, 0 }  ,fill opacity=1 ] (202.98,158.89) .. controls (202.98,158.33) and (203.43,157.89) .. (203.98,157.89) .. controls (204.53,157.89) and (204.98,158.33) .. (204.98,158.89) .. controls (204.98,159.44) and (204.53,159.89) .. (203.98,159.89) .. controls (203.43,159.89) and (202.98,159.44) .. (202.98,158.89) -- cycle ;
\draw  [fill={rgb, 255:red, 0; green, 0; blue, 0 }  ,fill opacity=1 ] (209.98,142.89) .. controls (209.98,142.33) and (210.43,141.89) .. (210.98,141.89) .. controls (211.53,141.89) and (211.98,142.33) .. (211.98,142.89) .. controls (211.98,143.44) and (211.53,143.89) .. (210.98,143.89) .. controls (210.43,143.89) and (209.98,143.44) .. (209.98,142.89) -- cycle ;
\draw  [fill={rgb, 255:red, 0; green, 0; blue, 0 }  ,fill opacity=1 ] (197.98,173.89) .. controls (197.98,173.33) and (198.43,172.89) .. (198.98,172.89) .. controls (199.53,172.89) and (199.98,173.33) .. (199.98,173.89) .. controls (199.98,174.44) and (199.53,174.89) .. (198.98,174.89) .. controls (198.43,174.89) and (197.98,174.44) .. (197.98,173.89) -- cycle ;
\draw  [fill={rgb, 255:red, 0; green, 0; blue, 0 }  ,fill opacity=1 ] (191.98,188.89) .. controls (191.98,188.33) and (192.43,187.89) .. (192.98,187.89) .. controls (193.53,187.89) and (193.98,188.33) .. (193.98,188.89) .. controls (193.98,189.44) and (193.53,189.89) .. (192.98,189.89) .. controls (192.43,189.89) and (191.98,189.44) .. (191.98,188.89) -- cycle ;

\draw   (95.02,89.32) .. controls (95.02,86.56) and (97.26,84.32) .. (100.02,84.32) .. controls (102.79,84.32) and (105.02,86.56) .. (105.02,89.32) .. controls (105.02,92.08) and (102.79,94.32) .. (100.02,94.32) .. controls (97.26,94.32) and (95.02,92.08) .. (95.02,89.32) -- cycle ; \draw   (96.49,85.78) -- (103.56,92.85) ; \draw   (103.56,85.78) -- (96.49,92.85) ;
\draw   (67.02,161.32) .. controls (67.02,158.56) and (69.26,156.32) .. (72.02,156.32) .. controls (74.79,156.32) and (77.02,158.56) .. (77.02,161.32) .. controls (77.02,164.08) and (74.79,166.32) .. (72.02,166.32) .. controls (69.26,166.32) and (67.02,164.08) .. (67.02,161.32) -- cycle ; \draw   (68.49,157.78) -- (75.56,164.85) ; \draw   (75.56,157.78) -- (68.49,164.85) ;
\draw   (188.02,84.32) .. controls (188.02,81.56) and (190.26,79.32) .. (193.02,79.32) .. controls (195.79,79.32) and (198.02,81.56) .. (198.02,84.32) .. controls (198.02,87.08) and (195.79,89.32) .. (193.02,89.32) .. controls (190.26,89.32) and (188.02,87.08) .. (188.02,84.32) -- cycle ; \draw   (189.49,80.78) -- (196.56,87.85) ; \draw   (196.56,80.78) -- (189.49,87.85) ;
\draw   (126.02,229.31) .. controls (126.02,226.55) and (128.26,224.31) .. (131.02,224.31) .. controls (133.79,224.31) and (136.02,226.55) .. (136.02,229.31) .. controls (136.02,232.07) and (133.79,234.31) .. (131.02,234.31) .. controls (128.26,234.31) and (126.02,232.07) .. (126.02,229.31) -- cycle ; \draw   (127.49,225.78) -- (134.56,232.85) ; \draw   (134.56,225.78) -- (127.49,232.85) ;
\draw    (100.02,89.32) .. controls (101.6,87.57) and (103.27,87.48) .. (105.02,89.05) .. controls (106.77,90.62) and (108.44,90.53) .. (110.01,88.78) .. controls (111.58,87.03) and (113.25,86.94) .. (115,88.51) .. controls (116.75,90.08) and (118.42,89.99) .. (120,88.24) .. controls (121.57,86.49) and (123.24,86.4) .. (124.99,87.98) .. controls (126.74,89.55) and (128.41,89.46) .. (129.98,87.71) .. controls (131.55,85.96) and (133.22,85.87) .. (134.97,87.44) .. controls (136.72,89.01) and (138.39,88.92) .. (139.97,87.17) .. controls (141.54,85.42) and (143.21,85.33) .. (144.96,86.9) .. controls (146.71,88.47) and (148.38,88.38) .. (149.95,86.63) .. controls (151.52,84.88) and (153.19,84.79) .. (154.94,86.37) .. controls (156.69,87.94) and (158.36,87.85) .. (159.94,86.1) .. controls (161.51,84.35) and (163.18,84.26) .. (164.93,85.83) .. controls (166.68,87.4) and (168.35,87.31) .. (169.92,85.56) .. controls (171.5,83.81) and (173.17,83.72) .. (174.92,85.29) .. controls (176.67,86.86) and (178.34,86.77) .. (179.91,85.02) .. controls (181.48,83.27) and (183.15,83.18) .. (184.9,84.75) .. controls (186.65,86.33) and (188.32,86.24) .. (189.89,84.49) -- (193.02,84.32) -- (193.02,84.32) ;
\draw    (100.02,89.32) .. controls (100.97,91.48) and (100.36,93.03) .. (98.2,93.97) .. controls (96.04,94.92) and (95.43,96.47) .. (96.37,98.63) .. controls (97.31,100.79) and (96.7,102.34) .. (94.54,103.28) .. controls (92.38,104.22) and (91.77,105.77) .. (92.72,107.93) .. controls (93.66,110.09) and (93.05,111.64) .. (90.89,112.59) .. controls (88.73,113.53) and (88.12,115.08) .. (89.06,117.24) .. controls (90,119.4) and (89.39,120.95) .. (87.23,121.9) .. controls (85.07,122.84) and (84.46,124.39) .. (85.41,126.55) .. controls (86.35,128.71) and (85.74,130.26) .. (83.58,131.21) .. controls (81.42,132.15) and (80.81,133.7) .. (81.75,135.86) .. controls (82.7,138.02) and (82.09,139.57) .. (79.93,140.51) .. controls (77.77,141.46) and (77.16,143.01) .. (78.1,145.17) .. controls (79.04,147.33) and (78.43,148.88) .. (76.27,149.82) .. controls (74.11,150.77) and (73.5,152.32) .. (74.44,154.48) .. controls (75.39,156.64) and (74.78,158.19) .. (72.62,159.13) -- (71.33,162.4) -- (71.33,162.4) ;
\draw    (71.33,162.4) .. controls (73.68,162.53) and (74.79,163.78) .. (74.66,166.13) .. controls (74.53,168.48) and (75.64,169.73) .. (77.99,169.86) .. controls (80.34,169.99) and (81.45,171.24) .. (81.32,173.59) .. controls (81.19,175.94) and (82.3,177.19) .. (84.65,177.32) .. controls (87,177.46) and (88.11,178.71) .. (87.98,181.06) .. controls (87.84,183.41) and (88.95,184.66) .. (91.3,184.79) .. controls (93.65,184.92) and (94.76,186.17) .. (94.63,188.52) .. controls (94.5,190.87) and (95.61,192.12) .. (97.96,192.25) .. controls (100.31,192.38) and (101.42,193.63) .. (101.29,195.98) .. controls (101.16,198.33) and (102.27,199.58) .. (104.62,199.71) .. controls (106.97,199.84) and (108.08,201.09) .. (107.95,203.44) .. controls (107.82,205.79) and (108.93,207.04) .. (111.28,207.17) .. controls (113.63,207.3) and (114.74,208.55) .. (114.6,210.9) .. controls (114.47,213.25) and (115.58,214.5) .. (117.93,214.64) .. controls (120.28,214.77) and (121.39,216.02) .. (121.26,218.37) .. controls (121.13,220.72) and (122.24,221.97) .. (124.59,222.1) .. controls (126.94,222.23) and (128.05,223.48) .. (127.92,225.83) -- (131.02,229.31) -- (131.02,229.31) ;
\draw    (193.02,84.32) .. controls (195.29,84.99) and (196.09,86.45) .. (195.42,88.71) .. controls (194.76,90.97) and (195.56,92.43) .. (197.82,93.09) .. controls (200.08,93.76) and (200.88,95.22) .. (200.22,97.48) .. controls (199.55,99.74) and (200.35,101.2) .. (202.61,101.87) .. controls (204.87,102.54) and (205.67,104) .. (205.01,106.26) .. controls (204.35,108.52) and (205.15,109.98) .. (207.41,110.64) -- (209.69,114.82) -- (209.69,114.82) ;
\draw    (131.02,229.31) .. controls (131.63,227.03) and (133.07,226.19) .. (135.35,226.8) .. controls (137.63,227.4) and (139.07,226.56) .. (139.67,224.28) .. controls (140.27,222) and (141.71,221.16) .. (143.99,221.77) .. controls (146.27,222.37) and (147.71,221.53) .. (148.31,219.25) .. controls (148.91,216.97) and (150.35,216.13) .. (152.63,216.73) .. controls (154.91,217.34) and (156.35,216.5) .. (156.95,214.22) .. controls (157.55,211.94) and (158.99,211.1) .. (161.27,211.7) .. controls (163.55,212.31) and (164.99,211.47) .. (165.59,209.19) -- (168.67,207.4) -- (168.67,207.4) ;

\draw (104.07,65.8) node [anchor=north west][inner sep=0.75pt]   [align=left] {$\displaystyle V_{1}$};
\draw (85.49,148.83) node [anchor=north west][inner sep=0.75pt]   [align=left] {$\displaystyle V_{2}$};
\draw (124.6,195.62) node [anchor=north west][inner sep=0.75pt]   [align=left] {$\displaystyle V_{3}$};
\draw (199.51,80.88) node [anchor=north west][inner sep=0.75pt]   [align=left] {$\displaystyle V_{N}$};

\end{tikzpicture}

\caption{\label{pics1} The defects on chiral strings. }
\end{figure}
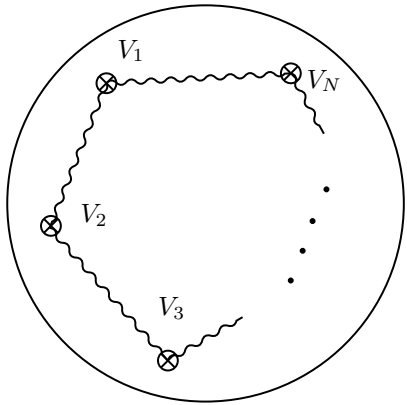

Therefore, we need to answer the questions of how to interpret the worldsheet model when the section condition is not satisfied. 
For general momentum, correlation functions $\langle\, V_{1}(z_{1})\cdots V_{n}(z_{n})\, \rangle$ are not analytic functions of vertex positions $z_{1}\cdots z_{n}$, but there are cuts on world-sheet which typically looks like figure \ref{pics1}. Those cuts could be interpreted as defects on world-sheet CFT. 

One consequence of defects is that now correlators are not completely symmetric in vertex operators' positions anymore. There are topologically inequivalent ordering of vertex operators on the edges of defects. In fact, correlators become only cyclic symmetric like open string correlators. 

\begin{figure}[h]
    \centering
    \caption{The three inequivalent configurations of 4-point amplitude. They roughly correspond to $s$, $t$, and $u$ channels of the amplitude.\label{pics2}}
    \import{.}{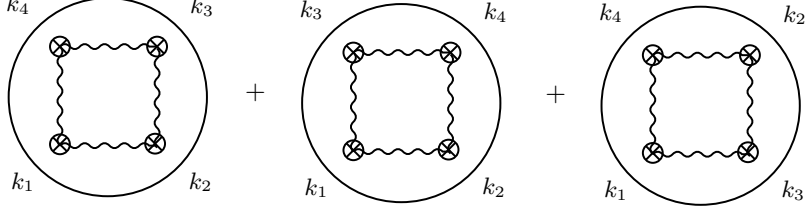}
\end{figure}

Therefore, in the calculation of amplitudes, one not only needs to integrate over vertex positions, but one also needs to sum over all topologically inequivalent vertex configurations. For example, there are three configurations that one needs to sum over in four-point amplitudes, see figure \ref{pics2}.

\FloatBarrier

As we will verify later by direct calculation, chiral string amplitudes are indeed only cyclic symmetric, and conditions for them to be completely symmetric are in fact \bref{sectioncond} and \bref{sectioncond2}.

We will only focus on tree amplitudes in this paper.  Amplitudes are given by standard prescription.
\[
M_{n}(1, \cdots, n) = \int_{\mathcal{M}}dm\,  \langle\, V(1), \cdots V(n) \,\rangle_{\Sigma}.
\]

Here, we use shorthand $V(i) = V_{i}(k_{iL}, k_{iR}; z_{i})$, and $\int_{\mathcal{M}}dm$ is integration over moduli space of correlators. The difference from standard string theory is that moduli space also needs to include topologically inequivalent configurations as mentioned above. Therefore, 
\[
\int_{\mathcal{M}}dm \equiv \sum_{\textrm{Defects topologies}}\int_{\mathbb{C}^{2}}d^{2}z_{2}\cdots\int_{\mathbb{C}^{2}}d^{2}z_{n-2}.
\]
 Chiral strings have the usual conformal Killing group that could fix 3 vertex positions. We fix $V_{1}, V_{n-1}, V_{n}$ to $0, 1, \infty$ respectively.  

To further evaluate the amplitudes, we will use similar procedures in deriving  KLT relations, where we analytically continue world-sheet coordinates $z,  \bar{z}$ to left and right moving coordinates $z_{+}, z_{-}$. 

\begin{equation}\label{acont}
\begin{matrix}
    z = \sigma_{1} + i\sigma_{0} \\ 
    \bar{z} = \sigma_{1} - i\sigma_{0} 
\end{matrix}, \quad \Rightarrow z_{\pm} =\sigma_{1} \pm \sigma_{0}   
\end{equation}

 Integrations over $z$ become a summation of integrations over various contours in $z_{+}, z_{-}$. Each term can be seen as the products of two open string amplitudes and a phase factor (Which become momentum kernels in KLT relations). \cite{Kawai:1985xq, Bjerrum-Bohr:2010pnr}.

There are some subtleties in doing this analytic continuation. Note that propagators listed in \bref{dprop} are oversimplified since we ignored $i\epsilon$ prescription. Let us start with the usual string propagators. (With both left and right moving modes included.)
\[
G(z, \bar{z})\sim \ln (z\bar z + i\epsilon) \quad \Rightarrow \quad \ln (z_{+}z_{-} + i\epsilon). 
\]

Of course, this is just the sum of left-moving and right-moving propagators, but because of the $i\epsilon$ term, the separation of expressions above to left-moving terms and right-moving terms are ambiguous. In particular,
\[
G(z_{+}, z_{-}) = G_{R}(z_{+}) + G_{L}(z_{-}) \sim \left \{
\begin{matrix}
    \ln\left(z_{+}\frac{z_{-} + i\epsilon}{|z_{-}|}\right) + \ln |z_{-}| \textrm{ , or} \\ 
    \ln |z_{+}| + \ln\left(z_{-}\frac{z_{+} + i\epsilon}{|z_{+}|}\right) 
\end{matrix} \right. .
\]

While this ambiguity has no effect when $k_{R} = k_{L}$, it does make a difference otherwise.
\begin{align}
& \left[k_{R}\cdot X_{R}(0) + k_{L}\cdot X_{L}(0)\right]\left[p_{R}\cdot X_{R}(z_{+}) + p_{L}\cdot X_{L}(z_{-})\right] \sim \nonumber\\
& k_{R}\cdot p_{R}\,\ln|z_{+}|+ k_{L}\cdot p_{L}\, \ln|z_{-}| +
\left \{
\begin{matrix}
  i\pi\, k_{R}\cdot p_{R}\,\theta(z_{+}z_{-})\\
  i\pi\, k_{L}\cdot p_{L}\,\theta(z_{+}z_{-})
\end{matrix} \right. \label{prop} ,
\end{align}
where $\theta(x)$ is the step function. Note that choices of separation of $G_{L}, G_{R}$ break inherent symmetry between left and right-moving modes, but such a choice just corresponds to exchanging left and right-moving modes. 

To calculate graviton amplitudes explicitly, we need to give a BRST prescription of the theory. That is, we need to introduce ghosts and construct integrated and unintegrated vertex operators $U(k_{R}, k_{L})$ and $W(k_{R}, k_{L}; z)$ respectively. As we shall see, at least for tree amplitudes, the BRST prescription is the same as usual strings.

$U(k_{R}, k_{L})$ and $W(k_{R}, k_{L}; z)$ for gravitons are 
\begin{align}
    & W(k_{R}, k_{L}; z) = c\bar{c}V(k_{R}, k_{L}; z), \quad U(k_{R}, k_{L}) = \int d^2{z} V(k_{R}, k_{L}; z)\nonumber \\
    & \quad V(k_{R}, k_{L}; z) = e_{\mu\nu}\pd X^{\mu}\bar{\pd}X^{\nu}e^{i k_{R}\cdot X_{R} + ik_{L}\cdot X_{L}}\label{gravexop}.
\end{align}

Where ghosts $c(z), \bar{c}(z)$ are same as in usual string theory\cite{Lize:2021una}. Using integrated and unintegrated vertex operators, one can define a gauge fix version of amplitude
\[
    M_{n}(1, \cdots, n) = \left\langle\, W(\hat{z}_{1}; K_{1})  U(K_{2})\cdots U(K_{n-2}) W(\hat{z}_{n-1}; K_{n-1}), W(\hat{z}_{n}; K_{n})\,\right\rangle,
\]
where $\hat{z}_{1}, \hat{z}_{n-1}, \hat{z}_{n}$ are positions that conformal Killing symmetry fixes. 

 Evaluation of amplitudes follows standard procedures, where we exponentiated  vertex operators

\[
    V(k_{R}, k_{L}; z) \rightarrow e^{ik_{R}\cdot X_{R} + ik_{L}\cdot X_{L}  + \varepsilon \cdot \pd X + \bar{\varepsilon} \cdot \bar{\pd} X.},
\]
and calculate correlators for exponentiated vertex operators, and extract multilinear terms in $\varepsilon $ and $\bar{\varepsilon}_{i}$ to get the original correlators. This avoids the need to compute complicated contractions between various fields.
After analytically continuing world-sheet coordinates \bref{acont}, amplitudes can be written as integration over left-moving coordinates $z_{-}$ and right-moving coordinates $z_{+}$.

\begin{align}\label{ampinte}
    M_{n} & \sim \int_{0}^{\infty} \prod_{i =2}^{n -2}dz_{i+} |z_{i+}|^{\frac{1}{2}k_{1R}\cdot k_{iR}}|1-z_{i+}|^{\frac{1}{2}k_{n-1R}\cdot k_{iR}}\prod_{i < j < n-1}|z_{ij+}|^{\frac{1}{2}k_{iR}\cdot k_{jR}}\;C(z_{i+}, k_{iR}, \varepsilon_{i})\nonumber\\
    \cdot & \int_{0}^{\infty} \prod_{i =2}^{n -2}dz_{i-} |z_{i-}|^{\frac{1}{2}k_{1L}\cdot k_{iL}}|1-z_{i-}|^{\frac{1}{2}k_{n-1L}\cdot k_{iL}}\prod_{i < j < n-1}|z_{ij-}|^{\frac{1}{2}k_{iL}\cdot k_{jL}}\;C(z_{i-}, k_{iL}, \bar{\varepsilon}_{i}) \nonumber\\
    \cdot  & \;\; \delta\left(\sum^{n}_{i}K_{i}\right) e^{i\pi \Phi(z_{1+}, z_{1-} \cdots z_{n-},  z_{n+})}.
\end{align}
Where functions $C(z_{i}, k_{i}, \varepsilon_{i})$ are rational functions in $z$ which comes from contractions between $\varepsilon_{i}\cdot\pd X$ and $e^{ik_{j}\cdot X}$ or $\varepsilon_{i}\cdot\pd X$ with $\varepsilon_{j}\cdot\pd X$, while Koba-Nielson factors comes from contractions between  $e^{ik_{i}\cdot X}$ and  $e^{ik_{j}\cdot X}$. 

We see in \bref{prop} that there is an extra term in propagators after analytic continuation, this gives us an additional phase factor $\Phi$ intertwining between the integration of left and right-moving coordinates. 
$$
    \Phi(z_{1+}, z_{1-} \cdots z_{n-},  z_{n+}) = \sum_{i < j}^{n}k_{iR}\cdot k_{jR} \, \theta (-z_{ij+}z_{ij-}).
$$
Recall that in \bref{prop} we have the freedom to choose how to decompose left moving and right moving modes due to $i\epsilon$ prescription, here we simply choose one that makes this phase factors proportional to $k_{iR}$, but there is no inherent difference if we choose the other decomposition.

In the integral expression of chiral string amplitude \bref{ampinte}, integrations over $z_{\pm}$ are just world-sheet integrations of open string amplitudes. Thus, graviton amplitudes could be written as the sum of bi-linear color-ordered open string amplitudes, intertwined with a phase (KLT kernel) due to the presence  $\Phi(z_{1+}, z_{1-} \cdots z_{n-},  z_{n+})$.

\begin{equation}\label{mklt}
    M_{n}(1, \cdots n) = \sum_{\sigma, \tau}\exp\left[i\pi  \sum_{i, j}^{n}k_{\sigma(i) R}\cdot  k_{\tau(j) R}\theta_{ij}(\sigma , \tau) \right]A^{(\textrm{L})}_{n}(\sigma(1, \cdots n ))A^{(\textrm{R})}_{n}(\tau(1, \cdots n)).
\end{equation}

$A^{(\textrm{L})}, A^{(\textrm{R})}$ are open string amplitudes, but their momenta are replaced by $k_{iL}, k_{iR}$ respectively. Summation over $\sigma, \tau$ are all permutations of $\{1, \cdots, n\}$ where the order of $1, n-1, n$ are fixed. Phase factor $\theta_{ij}(\sigma, \tau)$ is defined as
\begin{equation*}
    \theta_{ij}(\sigma , \tau) =    
    \begin{cases}
        1 & \text{If order of $\sigma(i), \sigma(j)$ and $\tau(i), \tau(j)$ are opposite.} \\
        0 & \text{If order of $\sigma(i), \sigma(j)$ and $\tau(i), \tau(j)$ are same.} 
    \end{cases}
\end{equation*}
This is the double copy representation of chiral string amplitude. While the complex phase factor seems unfamiliar, this is just the usual KLT momentum kernel introduced in \cite{Bjerrum-Bohr:2010pnr}.  This generalized KLT formula could be brought to a more familiar form if we use the fact that color-ordered open string amplitudes are not linearly independent. The $n!$ color-ordered amplitudes have only $(n-3)!$ independent components; they are related by monodromy relations \cite{Stieberger:2009hq,Bjerrum-Bohr:2009ulz}.

$$
    A(1, 2, \cdots , n ) + e^{i\pi k_{1}\cdot k_{2}}A(2, 1, \cdots, n) + \cdots  e^{i\pi k_{1}\cdot (k_{2}+\cdots k_{n-1})}A(2, \cdots, 1, n) = 0.
$$

We will work out explicit expressions for 4-point amplitudes.

\subsection{Four point amplitudes}\label{sec:Four-point-amplitudes}
For 4-point amplitude, we fix positions of vertex operators for $k_{1}, k_{2}$ and $k_{4}$. Doing so implicitly chooses topology for the defects. There are three independent ways one can choose fixed vertex operators, and they correspond to three different choices of defect topology in figure \ref{pics2}.

Using \bref{mklt} directly, the 4-point amplitude equals

\begin{align}\label{4ptraw}
    &M(1, 2, 3, 4) = \nonumber \\
    & A^{(\textrm{R})}(2, 1, 3, 4)\left[ A^{(\textrm{L})}(2, 1, 3, 4) +  e^{i\pi k_{1R}\cdot k_{2R}}A^{(\textrm{L})}( 1, 2, 3, 4)   +   e^{i\pi  k_{2R}\cdot( k_{1R} + k_{3R})}A^{(\textrm{L})}(1, 3, 2,  4) \right]\nonumber\\
    &A^{(\textrm{R})}(1, 2,  3, 4)\left[  e^{i\pi k_{1R}\cdot k_{2R}} A^{(\textrm{L})}(2, 1, 3, 4) + A^{(\textrm{L})}( 1, 2, 3, 4)   +   e^{i\pi k_{2R}\cdot k_{3R}}A^{(\textrm{L})}(1, 3, 2,  4) \right]\nonumber\\
    &A^{(\textrm{R})}(1, 3, 2, 4)\left[  e^{i\pi  k_{2R}\cdot (k_{1R} + k_{3R})} A^{(\textrm{L})}(2, 1, 3, 4) +e^{i\pi k_{2R}\cdot k_{3R}} A^{(\textrm{L})}( 1, 2, 3, 4)  + A^{(\textrm{L})}(1, 3, 2,  4) \right].\nonumber\\
\end{align}

Using monodromy relations, one could see that the first line and third line of expression above should vanish, while 
 second line could be simplified by subtracting a multiple of the following monodromy relation
\begin{equation}\label{mon}
A^{(\textrm{L})}(2, 1, 3, 4) +   e^{i\pi k_{1R}\cdot k_{2R}} A^{(\textrm{L})}( 1, 2, 3, 4)   +   e^{i\pi k_{2R}\cdot(k_{1R} + k_{3R})}A^{(\textrm{L})}(1, 3, 2,  4)  = 0.
\end{equation}
More specifically, 
\begin{align*}
    M(1, 2, 3, 4) =& A^{(\textrm{R})}(1, 2,  3, 4)\left[  e^{i\pi k_{1R}\cdot k_{2R}} A^{(\textrm{L})}(2, 1, 3, 4) + A^{(\textrm{L})}( 1, 2, 3, 4)   +   e^{i\pi k_{2R}\cdot k_{3R}}A^{(\textrm{L})}(1, 3, 2,  4) \right] \\
     -& A^{(\textrm{R})}(1, 2,  3, 4)\;  e^{-i\pi k_{1R}\cdot k_{2R}} \cdot \Bigl(\textrm{ LHS of equation \bref{mon}   }\Bigr)  \nonumber\\
    =& 2i\,A^{(\textrm{R})}(1, 2,  3, 4)\;\sin (\pi k_{1R}\cdot k_{2 R})\;A^{(\textrm{L})}(2, 1, 3, 4). \\
\end{align*}

Finally, adding the other two defect configurations, the full amplitude is (See Figure \ref{pics3}.)
\begin{align}
&M(1, 2, 3, 4) = \nonumber \\
&2i\left[\sin (\pi s_{R})A(s_{R}, u_{R})A(s_{L}, t_{L}) + \sin (\pi t_{R})A(t_{R}, s_{R})A(t_{L}, u_{L}) + \sin (\pi u_{R})A(u_{R}, t_{R})A(u_{L}, s_{L})\right].\label{4ptamp}
\end{align}
We have defined left and right-moving Mandelstam variables $s_{L, R}, t_{L, R}$ and $u_{L, R}$ similarly to usual Mandelstam variables. i.e.  $s_{R} = -(k_{1R} + k_{2R})^{2}$,  etc. It is not immediately obvious amplitude above is crossing symmetric, but if we write amplitude in terms of the gamma function, crossing symmetry becomes apparent.

\begin{equation}\label{4ptgamma}
M(1, 2, 3, 4) = {\cal K}_{L}\cdot{\cal K}_{R}\left(\frac{\Gamma(-s_{R})\Gamma(-t_{R})\Gamma(-u_{L})}{\Gamma( 1+ s_{L})\Gamma(1+ t_{L})\Gamma(1 + u_{R})} + (s \leftrightarrow u) + (t \leftrightarrow u)\right).
\end{equation}
This is the chiral string version of Virasoro amplitude. ${\cal K}_{L}$, ${\cal K}_{R}$ is just the kinematic factor of four-point open string amplitude, where momenta and polarization vectors are replaced by its left and right moving parts. Their expressions depend on what types of string theories they are, (e.g. bosonic or superstrings), and they are given explicitly at \cite{Siegel:2003sv}. In spacial case $k_{R} = k_{L}$,  ${\cal K}_{R}\cdot {\cal K}_{L}$ could be simplified to 4-graviton kinematic factor ${\cal W}^4$ mentioned in section 1. 
\begin{figure}[h]
    \centering
    \caption{The three different amplitudes of three different topological configurations of defects.\label{pics3}}
    \import{.}{figs3.tikz}
\end{figure}
\FloatBarrier

We can see that each term in \bref{4ptamp} is generally not $s, t, u$ symmetric, but there is still some symmetry implied by the topology of the defect. For example, the first piece of amplitude is symmetric under $t_{L, R} \leftrightarrow s_{L, R}$. This is consistent with the defect configuration shown in figure \ref{pics3}. While this symmetry is explicit if we write down amplitude in terms of Gamma functions, in order to generalize this result into higher order amplitude, is instructive to see how  $s \leftrightarrow t$ crossing manifests itself in KLT formula \bref{4ptamp}. 

By rewriting monodromy relations of 4-point amplitude differently, we have
$$
\sin (\pi s) A(s, t) = \sin (\pi u)A(u, t). 
$$
Since $A(s, t)$ is symmetric under $s, t$ exchange, we can see that $\sin (\pi s_{R})A(s_{R}, u_{R})A(s_{L}, t_{L}) $ is indeed symmetric under $s, t$ exchange.

\section{Section conditions}\label{sec:section-conditions}
When section condition \bref{sectioncond} holds, i.e. when every pair of momenta satisfy 
$$
k_{iR} \cdot k_{iR} - k_{jL} \cdot k_{jL}= 4n_{ij}, \quad n_{ij} \in \mathbb{Z},
$$
then are no defects, and we expect all three terms in \bref{4ptamp} equal to each other. This is the case.
By imposing \bref{sectioncond} means that $s_{L} = s_{R} + 4n_{12}$. If we substitute this equation into \bref{4ptgamma}, we will see that indeed all three terms will collapse into one single term. But it is more illuminating to verify $s, t, u$ crossing symmetry using monodromy relations of open strings directly.

When $s_{L}$ and $s_{R}$ differs by a multiple of 4,  $\sin(\pi s_{L}) = \sin(\pi s_{R})$. This means that we have the freedom to use monodromy relations for $A^{(R)}$ as well as $A^{(L)}$. To see how this works explicitly

\begin{align*}
    & \sin (\pi s_{R})A(s_{R}, u_{R})A(s_{L}, t_{L})   \\
    & =  \sin(\pi s_{L})A(s_{R}, u_{R})A(s_{L}, t_{L}) \quad\textrm{Apply $\sin (\pi s_{R}) \rightarrow \sin (\pi s_{L})$}.\\
    & =  \sin(\pi u_{L})A(s_{R}, u_{R})A(u_{L}, t_{L}) \quad\textrm{Apply monodromy relations on $A^{(L)}$.}\\
    & =  \sin(\pi u_{R})A(s_{R}, u_{R})A(u_{L}, t_{L}) \quad \textrm{Apply $\sin (\pi s_{L}) \rightarrow \sin (\pi s_{R})$. }
\end{align*}

Indeed, we find that the first term of \bref{4ptamp} equals the second term when there is no defect, and the fact that the first term and third term in \bref{4ptamp} are equal can be shown similarly. 

By verifying crossing symmetry in this way, we can see how we might generalize it to higher point amplitudes. Later we will show that any $n$-point amplitude is symmetric under permutations of external legs by applying monodromy relations repeatedly and using the fact that the momentum kernel doesn't change when replacing left-moving momenta to right-moving momenta (and vice versa).

 The derivation of crossing symmetry above relies on relation $\sin(\pi s_{L}) = \sin(\pi s_{R})$, but if we demand $\sin(\pi s_{L}) = -\sin(\pi s_{R})$ instead, derivation for crossing symmetry still holds. This corresponds to condition $s_{L} + s_{R} = 2n$ or twisted section condition \bref{sectioncond2}. Intriguingly, the twisted section condition still implies non-trivial monodromy around vertex operators, but different topological configurations on the worldsheet result in the same amplitude.
 
 As we shall see, under this twisted section condition, world-sheet CFT gives us a special kind of amplitudes which are rational functions of Mandelstam variables. Unlike string amplitudes, they contain a finite number of resonance poles, but usually, each resonance pole contains an infinite number of spins. The only exception for infinite spin towers is graviton amplitude.

\subsection{Solutions for section conditions}\label{sec:soulutions-for-section-conditions}
In the previous section, we find that when certain conditions are imposed on momenta of amplitudes. Amplitudes with different topological defect configurations are the same.

They are standard and twisted section conditions

\begin{align*}
&k_{iL}\cdot k_{jL} - k_{iR}\cdot k_{jR}  \equiv K_{i}^{M}\eta_{MN}K_{j}^{N} = 4n_{ij}, \quad n_{ij} \in \mathbb{Z}\\
&k_{iL}\cdot k_{jL} + k_{iR}\cdot k_{jR}  \equiv K_{i}^{M}\delta_{MN}K_{j}^{N} = 4m_{ij}, \quad m_{ij} \in \mathbb{Z}.
\end{align*}

For all pairs of external momenta $K_{i}, K_{j}$. Suppose that left or right-moving momenta lives in $d$ dimensional spacetime with $t$ time dimensions. $\eta_{N M}$ is the $SO(d, d)$ invariant tensor and $\delta_{NM}$ is the $SO(2t, 2d-2t)$ invariant tensor.
\[
\eta_{NM} = 
\begin{pmatrix}
-I_{t} & 0&  0 & 0\\
0 & I_{d-t}&  0 & 0\\
0 & 0&  I_{t} & 0\\
0 & 0&  0 & -I_{d-t}
\end{pmatrix}, \quad
\delta_{NM} = 
\begin{pmatrix}
 -I_{t}& 0&  0 & 0\\
0 & I_{d-t}&  0 & 0\\
0 & 0&   -I_{t}& 0\\
0 & 0&  0 &  I_{d-t}
\end{pmatrix}, 
\]
We shall try to find the most general solutions for external legs an amplitude could have under these section conditions.

We emphasize that ``momentum'' $K^{M}$ is just a parameter parameterizing the set of vertex operators in our theory, and it is not directly related to the physical momentum of the vertex operator. We aim to find a most general subset of momenta $S \subset \mathbb{R}^{d, d}$ that a consistent theory should have, and identify an embedding of physical momenta into $S$. 

There are a few conditions that $S$ needs to satisfy. First, any pair of momenta $K_{1}, K_{2}\in S$ need to satisfy standard section condition \bref{sectioncond} or twisted section condition \bref{sectioncond2}. Second, for any $K_{1}, K_{2} \in S$, $K_{1}+ K_{2}$ is in $S$. This is because OPEs of two vertex operators $V_{K_1}$ and $V_{K_2}$ contain $V_{K_{1}+ K_{2}}$. Third, $S$ contain a vector subspace $V$ of $\mathbb{R}^{d, d}$. In particular, for $K_{1}, K_{2}\in V$, $c_1K_{1} + c_{2}K_{2} \in V$, where $c_1, c_2$ are arbitrary real number. We can embed physical momenta into $V$. Thus, the dimension of $V$ is the dimension of chiral string theory.  

Let's solve $S$ for the standard section condition. $\mathbb{R}^{d, d}$ is endowed with standard $SO(d, d)$ invariant inner product $\langle \, K_{1} ,K_{2} \,\rangle = K_{1}^{M}\eta_{MN}K_{2}^{N}$.  Section condition is just the statement that the inner product of any two vectors in $S$ is multiple of 4.  Suppose that $v_{1}, v_{2} \in V$, since $c v_{2}$ are also in $V$ for any real number $c$, we have $c \langle\, v_{1}, v_{2}\, \rangle \in 4\mathbb{Z}$, or $ \langle\, v_{1}, v_{2}\, \rangle = 0$ for any  $v_{1}, v_{2} \in V$. i.e. $V$ needs to be a null subspace of $\mathbb{R}^{d, d}$.

Now suppose that some vector $v \in S$ but $v \notin V$. Decompose $v = v_{\perp} + v_{||}$ where $v_{\perp}$ is the component orthogonal of $V$  and $v_{||}$ is the component inside $V$. $\langle v, x\rangle = \langle v_{||}, x \rangle  \in 4\mathbb{Z}$ for all $x\in V$. Thus, $\langle v_{||}, x \rangle = 0$ or $v \in V_{\perp}$.

Combining the two results above shows that $S$ must be  
\begin{equation}\label{secsol}
 S = V \oplus \sum_{i}^{A} \mathbb{Z}e_{i}
\end{equation} 
where $V$ is some null subspace of $SO(d, d)$, and $e_{i} \in V_{\perp}$ is a set of vectors such that $\langle e_{i}, e_{j}\rangle \in 4\mathbb{Z}$.  The physical interpretation of this solution is clear.  Set of all possible vertex operator are labeled by $(\,\vec{p}\,, \,\vec{n}\,)$, where $\vec{p}\in V$ is physical momentum of vertex operator, while $\vec{n} = (n_{1}, \cdots n_{A}),\;\; n_{1}\cdots n_{M} \in \mathbb{Z}$ are  ``winding modes'' in a compactified direction.  $(\,\vec{p}\,, \,\vec{n}\,)$ are related to original momentum $K^{M}$ by 
\[
K^{M} = p^{M}  + \sum_{i}^{A}n_i\, e_{i}^{M} \equiv p^{M} + (n \cdot \mathbf{e})^{M} 
\]

With a suitable choice of basis in $\mathbb{R}^{d, d}$, any null vector in $p\in V$ could be written as
\[
p = (p{}',\; \underbrace{0 \cdots 0}_{d-k \textrm{ 0s}},\; p{}', \underbrace{0 \cdots 0}_{d-k \textrm{ 0s}}).
\]
$p{}'$ is a $k$-dimensional vector which is interpreted as the physical momentum of vertex operators. 

It's obvious that the set of consistent momenta $S$ for vertex operators are $SO(d, d; \mathbb{Z})$ invariant, and this formalism would construct manifestly $SO(d, d; \mathbb{Z})$ invariant amplitudes.

 The solution of $S$ for twisted section condition is essentially the same as \bref{secsol}, except that $V$ and $e_{i}$ are defined with a different inner product. Namely,  $\langle \, K_{1} ,K_{2} \,\rangle = K_{1}^{M}\delta_{MN}K_{2}^{N}$. If  chiral momenta lives in $\mathbb{R}^{d-t, t}$, $\delta_{NM}$ is just  invariant tensor for $SO(2t, 2d - 2t )$.  The maximum dimension of null space $V$ is the minimum of $2t$ and $2(d-t)$. In Lorentz spacetime, this means that $V$ could only be one-dimensional. But this is not a problem for us as we are dealing with tree amplitudes, and we could always Wick rotate momenta of Lorentzian space.

\section{Triple product amplitudes}\label{sec:triple-product-amplitudes}

In this section, we shall evaluate 4-point amplitudes under both standard and twisted section conditions explicitly, and check their various properties such as unitarity.

We will see that under these section conditions, the 4-point amplitude reduces to a triple product form $M(s, t, u) = F(s)F(t)F(u)$. These kinds of amplitudes are first discussed in \cite{Huang:2022mdb}, where they found a large number of triple-product amplitudes that are both unitary and UV-complete. Here we essentially found a world-sheet formulation for some of them.  

First, we consider standard section condition \bref{sectioncond}. As we have shown in the previous section. The momentum could be decomposed into

$$
K_{M} = (\,k_{m}, \bar{k}^{m}\,), \quad
\begin{cases}
     k_{m} = p_{m} + (n\cdot \mathbf{e})_{m}\\
     \bar{k}^{m} = \bar{p}^{m} + (\overline{n\cdot \mathbf{e}})^{m}\\
\end{cases},      
$$
where $n\cdot \mathbf{e} \equiv \sum n_{i}\vec{e}_{i}$ is a lattice point in $2d$-dimensional momentum space that specifies winding mode in compactified direction while $P_{M} = (p_{m}, \bar{p}^{m})$ is a $2d$ null vector that specify  momentum in  non-compactified direction. 

Let $K_{1M},\cdots K_{4M}$ be the vertex operator's momenta for external legs. Assume that all four of them have the same winding modes $\vec{n}$. The left and right-moving Mandelstam variables are
\begin{align}
    & s_{R} = -(k_{1m} + k_{2m})^{2} = -2p_{1}\cdot p_{2} - 4(n\cdot \mathbf{e})^{2}\nonumber\\
    & s_{L} = -(\bar{k}_{1}^{m} + \bar{k}_{2}^{m})^{2} = -2\bar{p}_{1}\cdot \bar{p}_{2} - 4(\overline{n\cdot \mathbf{e}})^{2}\label{msols}
\end{align} 
Let $p_{i} = \bar{p}_{i}$ be interpreted as actual physical momenta of the particles. $4(n\cdot \mathbf{e})^{2}$ and $4(\overline{n\cdot \mathbf{e}})^{2}$ are fixed integers which we will denote them as $r, \bar{r}$ respectively. Thus, we have 
$$s_{R} = s-r \textrm{ and, } s_{L} = s -\bar{r},$$
where $s \equiv -2p_{1}\cdot p_{2}$ are physical Mandelstam variables.  $t_{R}, t_{L}$ and $u_{R}, u_{L}$ are also related to $t, u$ by the same way. 

Substitute $s_{R, L}, t_{R, L} $ and $u_{R, L}$ into \bref{4ptgamma}, and upon simplification, we got 
$$
M(s, t, u) = {\cal K}_{R}\cdot{\cal K}_{L}\frac{\Gamma(-u + r)\Gamma(-t + r)\Gamma(-s +r)}{\Gamma(1+ u-\bar{r})\Gamma(1+ t-\bar{r})\Gamma(1+ s-\bar{r})}.
$$
To see the pole structure of this amplitude more easily, we could also write it in KLT factorized form
$$
M(s, t, u) = \sin \pi s_{R}\, A(s_{R}, t_{R})A(s_{L}, t_{L}) = \sin \pi s\, A(s-r, t-r)A(s-\bar{r}, u-\bar{r}).
$$

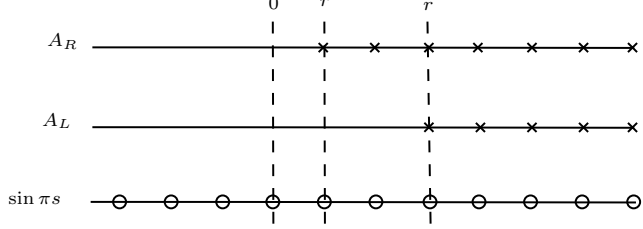
\begin{figure}[h]
    \centering
        \caption{How poles and zero cancel out for 4-point amplitude subjugated to standard section condition. Crosses are poles and circles are zeros.  \label{pics4}}
        \tikzset{every picture/.style={line width=0.75pt}} 

\begin{tikzpicture}[x=0.75pt,y=0.75pt,yscale=-1.3,xscale=1.3]

\draw    (100,110) -- (158.72,110.02) -- (310.22,110.22) ;
\draw    (100,141) -- (158.72,141.02) -- (310.22,141.22) ;
\draw    (99.33,170) -- (158.06,170.02) -- (310.89,170.22) ;
\draw  [dash pattern={on 4.5pt off 4.5pt}]  (170,100) -- (170.22,179.56) ;
\draw  [dash pattern={on 4.5pt off 4.5pt}]  (190,100) -- (190.22,179.22) ;
\draw  [dash pattern={on 4.5pt off 4.5pt}]  (230,100) -- (231.22,179.22) ;
\draw   (108.06,170) .. controls (108.06,168.62) and (109.17,167.5) .. (110.56,167.5) .. controls (111.94,167.5) and (113.06,168.62) .. (113.06,170) .. controls (113.06,171.38) and (111.94,172.5) .. (110.56,172.5) .. controls (109.17,172.5) and (108.06,171.38) .. (108.06,170) -- cycle ;
\draw   (128.06,170) .. controls (128.06,168.62) and (129.17,167.5) .. (130.56,167.5) .. controls (131.94,167.5) and (133.06,168.62) .. (133.06,170) .. controls (133.06,171.38) and (131.94,172.5) .. (130.56,172.5) .. controls (129.17,172.5) and (128.06,171.38) .. (128.06,170) -- cycle ;
\draw   (148.06,170) .. controls (148.06,168.62) and (149.17,167.5) .. (150.56,167.5) .. controls (151.94,167.5) and (153.06,168.62) .. (153.06,170) .. controls (153.06,171.38) and (151.94,172.5) .. (150.56,172.5) .. controls (149.17,172.5) and (148.06,171.38) .. (148.06,170) -- cycle ;
\draw   (167.5,170) .. controls (167.5,168.62) and (168.62,167.5) .. (170,167.5) .. controls (171.38,167.5) and (172.5,168.62) .. (172.5,170) .. controls (172.5,171.38) and (171.38,172.5) .. (170,172.5) .. controls (168.62,172.5) and (167.5,171.38) .. (167.5,170) -- cycle ;
\draw   (187.5,170) .. controls (187.5,168.62) and (188.62,167.5) .. (190,167.5) .. controls (191.38,167.5) and (192.5,168.62) .. (192.5,170) .. controls (192.5,171.38) and (191.38,172.5) .. (190,172.5) .. controls (188.62,172.5) and (187.5,171.38) .. (187.5,170) -- cycle ;
\draw   (207.5,170) .. controls (207.5,168.62) and (208.62,167.5) .. (210,167.5) .. controls (211.38,167.5) and (212.5,168.62) .. (212.5,170) .. controls (212.5,171.38) and (211.38,172.5) .. (210,172.5) .. controls (208.62,172.5) and (207.5,171.38) .. (207.5,170) -- cycle ;
\draw   (228.5,170) .. controls (228.5,168.62) and (229.62,167.5) .. (231,167.5) .. controls (232.38,167.5) and (233.5,168.62) .. (233.5,170) .. controls (233.5,171.38) and (232.38,172.5) .. (231,172.5) .. controls (229.62,172.5) and (228.5,171.38) .. (228.5,170) -- cycle ;
\draw   (247.5,170) .. controls (247.5,168.62) and (248.62,167.5) .. (250,167.5) .. controls (251.38,167.5) and (252.5,168.62) .. (252.5,170) .. controls (252.5,171.38) and (251.38,172.5) .. (250,172.5) .. controls (248.62,172.5) and (247.5,171.38) .. (247.5,170) -- cycle ;
\draw   (267.5,170) .. controls (267.5,168.62) and (268.62,167.5) .. (270,167.5) .. controls (271.38,167.5) and (272.5,168.62) .. (272.5,170) .. controls (272.5,171.38) and (271.38,172.5) .. (270,172.5) .. controls (268.62,172.5) and (267.5,171.38) .. (267.5,170) -- cycle ;
\draw   (287.5,170) .. controls (287.5,168.62) and (288.62,167.5) .. (290,167.5) .. controls (291.38,167.5) and (292.5,168.62) .. (292.5,170) .. controls (292.5,171.38) and (291.38,172.5) .. (290,172.5) .. controls (288.62,172.5) and (287.5,171.38) .. (287.5,170) -- cycle ;
\draw   (307.5,170) .. controls (307.5,168.62) and (308.62,167.5) .. (310,167.5) .. controls (311.38,167.5) and (312.5,168.62) .. (312.5,170) .. controls (312.5,171.38) and (311.38,172.5) .. (310,172.5) .. controls (308.62,172.5) and (307.5,171.38) .. (307.5,170) -- cycle ;
\draw   (228.79,139.27) -- (232.32,142.8)(232.32,139.27) -- (228.79,142.8) ;
\draw   (248.79,139.32) -- (252.32,142.85)(252.32,139.32) -- (248.79,142.85) ;
\draw   (268.79,139.32) -- (272.32,142.85)(272.32,139.32) -- (268.79,142.85) ;
\draw   (288.79,139.32) -- (292.32,142.85)(292.32,139.32) -- (288.79,142.85) ;
\draw   (307.79,139.32) -- (311.32,142.85)(311.32,139.32) -- (307.79,142.85) ;
\draw   (187.79,108.32) -- (191.32,111.85)(191.32,108.32) -- (187.79,111.85) ;
\draw   (207.79,108.32) -- (211.32,111.85)(211.32,108.32) -- (207.79,111.85) ;
\draw   (228.79,108.32) -- (232.32,111.85)(232.32,108.32) -- (228.79,111.85) ;
\draw   (247.79,108.32) -- (251.32,111.85)(251.32,108.32) -- (247.79,111.85) ;
\draw   (268.79,108.32) -- (272.32,111.85)(272.32,108.32) -- (268.79,111.85) ;
\draw   (288.79,108.32) -- (292.32,111.85)(292.32,108.32) -- (288.79,111.85) ;
\draw   (307.79,108.32) -- (311.32,111.85)(311.32,108.32) -- (307.79,111.85) ;

\draw (173, 93) node  [font=\scriptsize] [align=left] {\begin{minipage}[lt]{9.18pt}\setlength\topsep{0pt}
$\displaystyle 0$
\end{minipage}};
\draw (193, 93) node   [align=left] {\begin{minipage}[lt]{9.18pt}\setlength\topsep{0pt}
{\scriptsize $\displaystyle r$}
\end{minipage}};
\draw (233,93) node  [font=\scriptsize] [align=left] {\begin{minipage}[lt]{9pt}\setlength\topsep{0pt}
$ $$\displaystyle \overline{r}$
\end{minipage}};
\draw (88.97,107.56) node  [font=\scriptsize] [align=left] {\begin{minipage}[lt]{12.58pt}\setlength\topsep{0pt}
$\displaystyle A_{R}$
\end{minipage}};
\draw (86.97,138.56) node  [font=\scriptsize] [align=left] {\begin{minipage}[lt]{12.58pt}\setlength\topsep{0pt}
$\displaystyle A_{L}$
\end{minipage}};
\draw (73.97,168.56) node  [font=\scriptsize] [align=left] {\begin{minipage}[lt]{12.58pt}\setlength\topsep{0pt}
$\displaystyle \sin \pi s$
\end{minipage}};

\end{tikzpicture}
    \end{figure}

We see that the strings of poles coming from two open string amplitudes are shifted by $r$ and $\bar{r}$ respectively. Thanks to the sine factor, all potential double poles were reduced to a single poles. Figure \ref{pics4} displays more clearly how poles and zeros cancel each other, and shows the pole structure of final amplitude.

From figure \ref{pics4}, we can see that this amplitude has a string of resonance states with even spacing, where the lightest one starts at $\textrm{max}(r, \bar{r})$. 

Since $\Gamma(-x + \bar{r})/\Gamma( 1+ x -r) = (-1)^{r - \bar{r}}\Gamma(-x + r)/\Gamma( 1+ x - \bar{r}) $,  without loss of generality, we can assume that $\bar{r}\leq r$. We need $r = 0$ if we want to avoid tachyons and include massless gravitons. In this case, we could write the amplitude as 
\begin{equation}\label{sschrialamp}
M(s, t, u) = {\cal K}_{R}\cdot{\cal K}_{L}\frac{\Gamma(-u )\Gamma(-t )\Gamma(-s)}{\Gamma(1+ u-\bar{r})\Gamma(1+ t-\bar{r})\Gamma(1+ s-\bar{r})} = \frac{M_{\textrm{grav}}(s, t,u)}{(1 + s)_{-\bar{r}}(1 + t)_{-\bar{r}}(1 + u)_{-\bar{r}}},
\end{equation} where $M_{\textrm{grav}}(s, t,u)$ is standard 4-graviton amplitude.

Unfortunately, the fact that in this case, chiral string amplitude equals graviton amplitude times a rational function of $s, t, u$ means that its residues on resonance poles will not be a polynomial of  $x \equiv \cos \theta$, where $\theta$ is scattering angle. That means some resonance channels have an infinite tower of spin exchanges,  but as we shall see, only a finite number of such channels are non-polynomial of $x$. 

Now we consider the case for twisted section conditions \bref{sectioncond2}. We have the same situation as \bref{msols} for four external vertex operator's momenta,  except now $\bar{p}_{i} = -p_{i}$. We also redefined $4(n\cdot \mathbf{e})^{2}$ and $4(\overline{n\cdot \mathbf{e}})^{2}$ to be  $r, -\bar{r}$ respectively. Left-moving and right-moving Mandelstam variables are
$$
s_{R} = s-r \textrm{ and, } s_{L} = -s +\bar{r}.
$$
All other Mandelstam variables are defined the same way.

Upon simplification, the 4-point chiral string amplitude is 
$$
M(s, t, u) = {\cal K}_{R}\cdot {\cal K}_{L}\frac{\Gamma(u - \bar{r})\Gamma(t - \bar{r})\Gamma(s -\bar{r})}{\Gamma(1+ u - r)\Gamma(1+ t-r)\Gamma(1+ s-r)}.
$$
This expression is always a rational function of Mandelstam variables. 
\begin{align}
&M_{\textrm{Chiral}} = {\cal K}_{R}\cdot {\cal K}_{L}\nonumber\\
&   \begin{cases}
     { \scriptstyle (1+u -r)_{r - \bar{r} - 1}(1+t -r)_{r - \bar{r} - 1}(1+s-r)_{r - \bar{r} - 1}}& \text{, $r - \bar{r} - 1 \geq 0$}\\
      \frac{1}{( u -\bar{r})_{\bar{r}- r + 1}} \frac{1}{( t-\bar{r})_{\bar{r}- r + 1}} \frac{1}{( s -\bar{r})_{\bar{r}- r + 1}} & \text{,  $r - \bar{r} - 1 < 0$}
    \end{cases}\label{twstedamp}       
\end{align}
One can see that in some cases this amplitude is a polynomial in Mandelstam variables, which means interactions are higher derivative contact terms. These kinds of amplitudes are necessarily not UV-complete. In other cases, the amplitude has an infinite tower of higher spin exchanges in its resonance poles. 

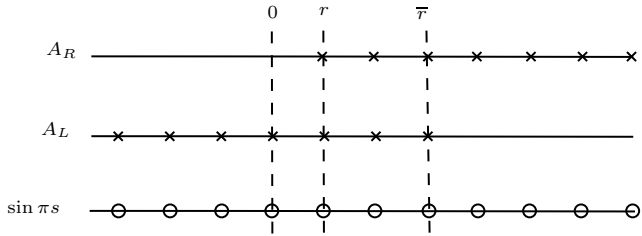
\begin{figure}[h]
    \centering
        \caption{How poles and zero cancel out for 4-point amplitude subjugated to twisted section condition.  Crosses are poles and circles are zeros.  \label{pics5}}
        \tikzset{every picture/.style={line width=0.75pt}} 

\begin{tikzpicture}[x=0.75pt,y=0.75pt,yscale=-1.3,xscale=1.3]

\draw    (100,110) -- (158.72,110.02) -- (310.22,110.22) ;
\draw    (100,141) -- (158.72,141.02) -- (310.22,141.22) ;
\draw    (99.33,170) -- (158.06,170.02) -- (310.89,170.22) ;
\draw  [dash pattern={on 4.5pt off 4.5pt}]  (170,100.33) -- (170.22,179.56) ;
\draw  [dash pattern={on 4.5pt off 4.5pt}]  (190,100) -- (190.22,179.22) ;
\draw  [dash pattern={on 4.5pt off 4.5pt}]  (230,100) -- (231.22,179.22) ;
\draw   (108.06,170) .. controls (108.06,168.62) and (109.17,167.5) .. (110.56,167.5) .. controls (111.94,167.5) and (113.06,168.62) .. (113.06,170) .. controls (113.06,171.38) and (111.94,172.5) .. (110.56,172.5) .. controls (109.17,172.5) and (108.06,171.38) .. (108.06,170) -- cycle ;
\draw   (128.06,170) .. controls (128.06,168.62) and (129.17,167.5) .. (130.56,167.5) .. controls (131.94,167.5) and (133.06,168.62) .. (133.06,170) .. controls (133.06,171.38) and (131.94,172.5) .. (130.56,172.5) .. controls (129.17,172.5) and (128.06,171.38) .. (128.06,170) -- cycle ;
\draw   (148.06,170) .. controls (148.06,168.62) and (149.17,167.5) .. (150.56,167.5) .. controls (151.94,167.5) and (153.06,168.62) .. (153.06,170) .. controls (153.06,171.38) and (151.94,172.5) .. (150.56,172.5) .. controls (149.17,172.5) and (148.06,171.38) .. (148.06,170) -- cycle ;
\draw   (167.5,170) .. controls (167.5,168.62) and (168.62,167.5) .. (170,167.5) .. controls (171.38,167.5) and (172.5,168.62) .. (172.5,170) .. controls (172.5,171.38) and (171.38,172.5) .. (170,172.5) .. controls (168.62,172.5) and (167.5,171.38) .. (167.5,170) -- cycle ;
\draw   (187.5,170) .. controls (187.5,168.62) and (188.62,167.5) .. (190,167.5) .. controls (191.38,167.5) and (192.5,168.62) .. (192.5,170) .. controls (192.5,171.38) and (191.38,172.5) .. (190,172.5) .. controls (188.62,172.5) and (187.5,171.38) .. (187.5,170) -- cycle ;
\draw   (207.5,170) .. controls (207.5,168.62) and (208.62,167.5) .. (210,167.5) .. controls (211.38,167.5) and (212.5,168.62) .. (212.5,170) .. controls (212.5,171.38) and (211.38,172.5) .. (210,172.5) .. controls (208.62,172.5) and (207.5,171.38) .. (207.5,170) -- cycle ;
\draw   (228.5,170) .. controls (228.5,168.62) and (229.62,167.5) .. (231,167.5) .. controls (232.38,167.5) and (233.5,168.62) .. (233.5,170) .. controls (233.5,171.38) and (232.38,172.5) .. (231,172.5) .. controls (229.62,172.5) and (228.5,171.38) .. (228.5,170) -- cycle ;
\draw   (247.5,170) .. controls (247.5,168.62) and (248.62,167.5) .. (250,167.5) .. controls (251.38,167.5) and (252.5,168.62) .. (252.5,170) .. controls (252.5,171.38) and (251.38,172.5) .. (250,172.5) .. controls (248.62,172.5) and (247.5,171.38) .. (247.5,170) -- cycle ;
\draw   (267.5,170) .. controls (267.5,168.62) and (268.62,167.5) .. (270,167.5) .. controls (271.38,167.5) and (272.5,168.62) .. (272.5,170) .. controls (272.5,171.38) and (271.38,172.5) .. (270,172.5) .. controls (268.62,172.5) and (267.5,171.38) .. (267.5,170) -- cycle ;
\draw   (287.5,170) .. controls (287.5,168.62) and (288.62,167.5) .. (290,167.5) .. controls (291.38,167.5) and (292.5,168.62) .. (292.5,170) .. controls (292.5,171.38) and (291.38,172.5) .. (290,172.5) .. controls (288.62,172.5) and (287.5,171.38) .. (287.5,170) -- cycle ;
\draw   (307.5,170) .. controls (307.5,168.62) and (308.62,167.5) .. (310,167.5) .. controls (311.38,167.5) and (312.5,168.62) .. (312.5,170) .. controls (312.5,171.38) and (311.38,172.5) .. (310,172.5) .. controls (308.62,172.5) and (307.5,171.38) .. (307.5,170) -- cycle ;
\draw   (228.79,139.27) -- (232.32,142.8)(232.32,139.27) -- (228.79,142.8) ;
\draw   (187.79,108.32) -- (191.32,111.85)(191.32,108.32) -- (187.79,111.85) ;
\draw   (207.79,108.32) -- (211.32,111.85)(211.32,108.32) -- (207.79,111.85) ;
\draw   (228.79,108.32) -- (232.32,111.85)(232.32,108.32) -- (228.79,111.85) ;
\draw   (247.79,108.32) -- (251.32,111.85)(251.32,108.32) -- (247.79,111.85) ;
\draw   (268.79,108.32) -- (272.32,111.85)(272.32,108.32) -- (268.79,111.85) ;
\draw   (288.79,108.32) -- (292.32,111.85)(292.32,108.32) -- (288.79,111.85) ;
\draw   (307.79,108.32) -- (311.32,111.85)(311.32,108.32) -- (307.79,111.85) ;
\draw   (208.64,139.27) -- (212.18,142.8)(212.18,139.27) -- (208.64,142.8) ;
\draw   (188.64,139.27) -- (192.18,142.8)(192.18,139.27) -- (188.64,142.8) ;
\draw   (168.64,139.27) -- (172.18,142.8)(172.18,139.27) -- (168.64,142.8) ;
\draw   (148.64,139.27) -- (152.18,142.8)(152.18,139.27) -- (148.64,142.8) ;
\draw   (128.64,139.27) -- (132.18,142.8)(132.18,139.27) -- (128.64,142.8) ;
\draw   (108.64,139.27) -- (112.18,142.8)(112.18,139.27) -- (108.64,142.8) ;

\draw (173,93) node  [font=\scriptsize] [align=left] {\begin{minipage}[lt]{9.18pt}\setlength\topsep{0pt}
$\displaystyle 0$
\end{minipage}};
\draw (193,93) node   [align=left] {\begin{minipage}[lt]{9.18pt}\setlength\topsep{0pt}
{\scriptsize $\displaystyle r$}
\end{minipage}};
\draw (233,93) node  [font=\scriptsize] [align=left] {\begin{minipage}[lt]{13.03pt}\setlength\topsep{0pt}
$ $$\displaystyle \overline{r}$
\end{minipage}};
\draw (88.97,107.56) node  [font=\scriptsize] [align=left] {\begin{minipage}[lt]{12.58pt}\setlength\topsep{0pt}
$\displaystyle A_{R}$
\end{minipage}};
\draw (86.97,138.56) node  [font=\scriptsize] [align=left] {\begin{minipage}[lt]{12.58pt}\setlength\topsep{0pt}
$\displaystyle A_{L}$
\end{minipage}};
\draw (73.97,168.56) node  [font=\scriptsize] [align=left] {\begin{minipage}[lt]{12.58pt}\setlength\topsep{0pt}
$\displaystyle \sin \pi s$
\end{minipage}};

\end{tikzpicture}
\end{figure}

The only exception in both cases is when $r = \bar{r} = 0$ where the chiral string is reduced to four graviton amplitude.     
One can also write this amplitude as a double copy of open string amplitudes 
$$
M(s, t, u) = \sin \pi s\, A(s-r, t-r)A(-s+\bar{r}, -u+\bar{r}).
$$ 
Written in this form, we could see how double pole cancellation works. Unlike the untwisted case, there is only a finite number of poles. (See picture \ref{pics5}.) 

We see that in cases when $r  \leq \bar{r}$, there is only a finite string of poles starting from $s = r$  and ending at $s = \bar{r}$. Thus, to avoid tachyons and include massless states, $r$ is fixed to 0. 

\subsection{Unitarity constraints}\label{sec:unitarity-constraints}
In this section, we would like to study the unitarity of chiral string amplitudes. 
In particular, we would like to focus on the triple product amplitudes we found in the previous section.

For amplitudes to be unitary, one of the necessary conditions is for their residues of resonance poles to admit a positive expansion of Gegenbauer polynomials.\footnote{This is not sufficient condition though, additional conditions are coming from consistent factorization for higher point amplitudes\cite{Arkani-Hamed:2023jwn}.} More precisely, given a 4pt amplitude $A(s, t)$ which has a $s$-channel resonance poles at $M^{2}$. Its residues denoted as $R_{M}(t) = -\textrm{Res}{}_{s = M^{2}}A(s, t)$ could be expanded in powers of $x = \cos \theta$, where $\theta$ is scattering angle, and $t = -M^{2}(1-x)/2$.

We could expand $R_{M}(x)$ in terms of Gegenbauer polynomials $C^{(\alpha)}_{\ell}(x)$.
$$
R_{M}(x) = \sum_{\ell = 0}^{N}a_{\ell}C^{(\alpha)}_{\ell}(x).
$$
One of the conditions for the amplitude to be unitary is that $a_{\ell} \geq 0$. In this case we call $R_{M}(x)$ positive \cite{Arkani-Hamed:2022gsa}.
$\alpha = \frac{(d - 3)}{2}$, $d$ is the dimensionality of the theory. This means unitarity constraint also depends on the dimension. All individual terms in the expansion of $R_{M}(x)$ could be understood as a spin-$\ell$ particle exchange, and $a_{\ell}$ is the square of 2-point coupling constants of external particles with  spin-$\ell$ exchanging particles.

Generally, $R_{M}(x)$ is a polynomial which means that there is a maximum spin exchange,  but as we saw in previous sections, $R_{M}$ might not be a polynomial, and in this case, the theory has a tower of arbitrary high spin exchanges at a given mass level $M$. This type of behavior appears in solutions of the EFT bootstrap program. See \cite{Albert:2022oes, Berman:2023jys} for the infinite spin tower in spin-1 amplitudes bootstrap  \cite{Albert:2024yap} for gravitational amplitude bootstrap.

Consider the triple-product amplitude   
$$
M(s, t, u) = F(s)F(t)F(u).
$$
Assuming that $F(s)$ is a meromorphic function with poles at $s = m^2_{a}$, which correspond to masses of resonance states. We also assume that there is a massless exchange $m_{0}^{2} = 0$. Let $g_{a} = -\textrm{Res}_{s = m^2_a}F(s)$, and normalization is defined as such that $g_{0} = -1$. 

Formally, we have an expansion
$$
F(s) = \frac{1}{s} + \sum_{a= 1}\frac{g_{a}}{-s + m_{a}^2}.
$$
Residue on $m_{a}^2$ channel is 
$$
R_{a}(x) \equiv\textrm{Res}_{s = m^2_a}M(s, t, u) = g_{a}F\left(-m^2_a\frac{1+x}{2}\right)F\left(-m^2_a\frac{1-x}{2}\right).
$$
It is shown in \cite{Huang:2022mdb} that if $g_{a} \geq 0$, then $R_{a}(x)$ is positive. Of course, there are extra checks for triple product amplitude such as it needs to satisfy Regge bound and UV-softness \cite{Chowdhury:2019kaq, Haring:2022cyf}.
That is 
$$
\lim_{s\rightarrow\infty} \frac{M(s, t)}{|s|^2} = 0, \quad \textrm{for $t < 0$},
$$
and UV-completeness simply means that the amplitude won't diverges in the fix scattering angle, large energy limit. 
These constraints will give further restriction on $g_{a}$, and we will also check them for chiral string amplitudes.

First, we shall focus on the unitarity properties of chiral string amplitudes under standard section conditions. We only consider the case when there is a massless exchange. i.e.  Amplitude of the form \bref{sschrialamp}.

In this case, $F(s)$ equals  
\begin{equation}\label{standardF}
F(s) = \frac{1}{(1+s)_{k}}\frac{\Gamma(-s)}{\Gamma(1+s)}.
\end{equation}
For simplicity, we have redefined $-\bar{r} = k$, where $k$ should be non-negative integers.
Poles of $F(s)$ are located at $s = n$, and their residues are
$$
g_{n} \equiv -\textrm{Res}_{s = n}F(s) = \frac{(-1)^{n}}{(n!)^{2}(1 + n)_{k}}.
$$
Residues are not all positive, so we cannot apply directly conclusion in the original paper \cite{Huang:2022mdb}, and we have to calculate $R_{n}(x) = -\textrm{Res}_{s=n}M(s, t, u)$ directly.
\begin{align*}
&R^{(k)}_{n}(x) = g_{n}F\left(-n\frac{1+x}{2}\right)F\left(-n\frac{1-x}{2}\right)\\
& = (-1)^{k-1}\frac{\displaystyle\left(\frac{n}{2}\right)^{2n -2 + 2k}}{(n!)^{2}(1+n)_{k}}\frac{\displaystyle\prod_{i = 1}^{n-1}\left[x^{2}  -\left(\frac{n-2i}{n}\right)^{2}\right]}{\displaystyle \prod_{i = 1}^{k}\left[x^{2}  -\left(\frac{n-2i}{n}\right)^{2}\right]}
\end{align*}
If $n > k$ then the residue is a polynomial, otherwise, it's a rational function. Thus, for any given $k$ only a finite number of poles have arbitrarily large spin exchanges. 

In other words,
$$
R^{(k)}_{n}(x) = 
\begin{cases}
\begin{aligned}
    {\displaystyle(-1)^{k-1} c_{n, k}\prod_{i = k+1}^{n-1}\left[x^{2}  -\left(\frac{n-2i}{n}\right)^{2}\right]}& \;\; n > k\\[1ex]
    {\displaystyle (-1)^{k-1} c_{n, k}\prod_{i = n}^{k}\left[x^{2}  -\left(\frac{n-2i}{n}\right)^{2}\right]^{-1}}& \;\; n \leq k
\end{aligned}
\end{cases},
$$
where $c_{n, k} = (n/2)^{2n+ 2k + 2}/((n!)^{2}(1+n)_{k}) > 0$.

Let's give a few specific examples for standard cases. If $k = 1$. $R_{n}(x)$ is a polynomial for $n \geq 2$. In particular, we have 
\begin{align*}
    R^{(1)}_{2}(x) & = \frac{1}{12} = \frac{1}{12}G_{0}^{(\alpha)}(x)\\
    R^{(1)}_{3}(x) & = \frac{x^{2}}{64} - \frac{1}{576} = \frac{1}{(d-3)(d-1)}G^{(\alpha)}_{2}(x) + \frac{10-d}{576(d-1)}G^{\alpha}_{0}(x)\\
    R^{(1)}_{4}(x) & = \frac{x^{4}}{180} - \frac{x^2}{720}= \frac{2}{15(d^{2}-1)(d^{2}-9)}G_{4}^{(\alpha)}(x)+\frac{21-d}{360(d-1)(d^{2}-9)}G_{2}^{(\alpha)}(x)\\
    &+\frac{11-d}{15(d^{2}-1)}G_{0}^{(\alpha)}(x)
\end{align*}
We can see that for polynomials to be positive (and well-defined), we need $4 \leq d \leq 10$. A further numerical evaluation up to $n \leq 30$ does not impose extra conditions on dimension $d$. 

For $k =2$, a similar analysis showed that $R^{(2)}_{n}(x)$ seem to be \emph{pure negative} when $d \leq 5$ (Explicitly verified for $n < 30$). For $d > 5$, they contain both positive and negative terms. The pure negativity is due to factor $(-1)^{k-1}$ in front of $R^{k}_{n}(x)$. 

For $k \leq 3$, $R^{(3)}_{n}(x)$ could be both positive and negative even when $d = 4$. Most constraining residues seems to be $R^{(k)}_{k+2}(x)$. 

$$
R^{(k)}_{k+2}(x)=\frac{(-1)^{k-1}}{(k+2)!(2k+2)!}\left(\frac{(2-d)k^{2} + 4k + 4}{4(d-1)}G^{(\alpha)}_{0}(x)+ \frac{(k+2)^{3}}{2(d-3)(d-1)}G_{2}^{(\alpha)}(x)\right)
$$
One can see that terms proportional to $G^{(\alpha)}_{0}(x)$ and $G^{(\alpha)}_{2}(x)$ could be same sign if and only if $d \leq 2\frac{k^{2} + 2k + 2}{k^{2}}$. This value becomes less than $4$ when $k \geq 3$.

We should also analyze non-polynomial residues for $k = 1, 2$. These residues are
$$
R^{(1)}_{1}(x)= -\frac{2}{1-x^{2}}, \quad R^{(2)}_{1}(x)= -\frac{8}{3(1-x^{2})(9-x^{2})},\quad R^{(2)}_{2}(x)= \frac{1}{48(1-x^{2})}.
$$

Expand these functions in terms of $x$. $R_{1}^{(1)}(x)$ and $ R^{(2)}_{1}(x)$ is a pure negative function, while $R_{2}^{(2)}(x)$ is a pure positive function.\footnote{To see this, monomials $x^{k}$ is a positive expansion of Gegenbauer polynomials, and thus $\frac{1}{1-x^{2}} = \sum_{k=0}^{\infty}x^{k}$ is also positive. Also, products of positive polynomials are themselves positive, thus $\frac{1}{(9-x^{2})(1-x^{2})}$ is positive.} This rules out amplitudes to be unitary for $k= 1, 2$ because, for $k =1$, polynomial residues are all positive and thus have different signs with non-polynomial residues.

Unfortunately, for standard section conditions, we didn't find any amplitudes that satisfy unitarity constraints other than ordinary Virasoro amplitude.
But as we shall discuss later, the analysis above is not a comprehensive search of all possible amplitude. 

Next, we shall study the unitarity properties of chiral string amplitude under twisted section conditions. We will only consider amplitudes with the form \bref{twstedamp} where $r = 0$ and $k\equiv \bar{r} \geq 0$. That is when amplitudes are triple products of rational functions. 

This amplitude has poles at $ s = 0, 1, \cdots k$. 
\begin{equation}\label{twistedF}
F(s) = \frac{1}{(s - k)_{k + 1}} = \frac{1}{s(s-1)\cdots(s-k)}.
\end{equation}
Residues at its poles are 
$$
g_{n} = -\mathop{\textrm{Res}}_{s = n}F(s) = \frac{(-1)^{k + n +1}}{(k - n)!n!}
$$
Corresponding residues $\tilde{R}^{(k)}_{n}(x)$ is 
\begin{align*}
&\tilde{R}^{(k)}_{n}(x) = (-1)^{n-1}\frac{(n/2)^{2k +2}}{n!(k - n)!} \prod_{i = 0}^{k}\left[x^{2}  -\left(\frac{n+2i}{n}\right)^{2}\right]^{-1}
\end{align*}
As expected, residues are always rational functions of $x$. 

The factor $\prod_{i = 0}^{k}\left[x^{2}  -\left(\frac{n+2i}{n}\right)^{2}\right]^{-1}$ is always a positive expansion of Gegenbauer polynomial. Therefore, residues are pure positive when $n$ is odd and pure negative when $n$ is even. 
The only possible way for all residues to remain one sign is when there is only a single channel. i.e. when $k = 1$.  

In fact, triple product factor $F(s)$ could be rewritten as  
$$
F(s) = \frac{1}{s(s-1)} = \frac{1}{s} + \frac{1}{1-s}.
$$
This is a specific case proposed in the original paper on triple product amplitudes \cite{Huang:2022mdb}. Intriguingly, this type of amplitude (Amplitude that contains one massless pole with finite spin plus one massive pole with infinite spin tower) appeared in amplitude bootstrap programs as kinks on allowed spaces of Wilson coefficients \cite{Albert:2022oes, Albert:2024yap, Berman:2023jys}.

From the preliminary analysis of the previous section, the only non-trivial unitary amplitude we find is 
\begin{equation}\label{ampist}
M(s, t, u) = {\cal K}_{R}\cdot {\cal K}_{L} F(s)F(t)F(u), \quad F(s) = \frac{1}{s}  + \frac{1}{1-s}.
\end{equation}

Some brief comments on Regge bound and UV completeness of chiral string amplitudes.  For amplitudes with standard section condition \bref{sschrialamp}, they have Regge limit $M(s, t, u) \sim t^{-s - 2 - 2k}$. The slope of Regge trajectories is the same as string theory, but intercepts are controlled by positive integer $k$. 

Of course, the fact that this amplitude is a product of Virasoro amplitude and rational functions of $s, t, u$, means that in fixed angle, large energy limit it is exponentially soft and thus UV-complete.

For amplitudes with twisted section condition \bref{twstedamp}, Regge trajectories are  $M(s, t, u) \sim t^{- 2 - 2k}$, and its fixed angle, large energy limit goes as inverse powers of energy $ (E)^{2 - 6k}$. In particular, unitary amplitude \bref{ampist} decays as $E^{-4}$, which satisfy Regge bound and UV-completeness \cite{Chowdhury:2019kaq}.

\subsection{More general amplitudes}\label{sec:more-general-amplitudes}
In the previous section, we find one non-trivial unitary amplitude candidate \bref{ampist}. One might think of possible generalizations or deformations of chiral string amplitudes.

If we cast aside the worldsheet construction of these amplitudes, and just focus on satisfying unitarity, Regge bound, and UV-completeness of amplitudes, then there is a straightforward deformation of chiral string amplitudes. We can always multiply by some polynomials of Mandelstam variables $P(s, t, u)$.

\begin{equation}\label{defromedchamp}
 M_{k}(s, t, u) =  {\cal W}^4 P(s, t, u)F_{k}(s)F_{k}(t)F_{k}(u) 
\end{equation}
Of course, this polynomial needs to satisfy some constraints. For this amplitude to be crossing symmetric, $P(s, t,u)$ is a symmetric polynomial of its variables. 
 A way to keep amplitude unitary is for $P(s, t, u)$ to be positive after substituting $s, t, u$ with scattering angle, i.e. $P(M^{2}, -M^2(1+x)/2, -M^2(1-x)/2)$ is a positive polynomial of $x$ for all resonance states $s = M^2$. (Since products of positive polynomials are still positive.), and to satisfy Regge bound and UV-completeness, the total degree of $P(s, t, u)$ needs to be less than $3k$.

As an example, let's consider the deformation of amplitude under twisted section condition \bref{twstedamp}. We restricted ourselves to a special case where $P$ itself is in triple product form $P(s, t, u) = P(s)P(t)P(u)$. In general, we can always write (up to an overall scale)
$$
P(s)F_{k}(s) =\frac{P(s)}{s(s-1)\cdots(s-k)} = \frac{1}{s}  + \frac{c_1}{1-s} +\cdots + \frac{c_k}{k-s},
$$ 
where constants $c_1, \cdots c_{k}$ could be fine-tuned. Thus, if we set $c_{i} > 0$ and $\sum_{i} c_{i} = 1$, we arrive at the same triple product amplitudes described in \cite{Huang:2022mdb}.

It is natural to ask whether the proposed deformed chiral string amplitude \bref{defromedchamp} could have a button-up derivation from worldsheet theories. While we do not have a definitive answer to it yet, we would like to propose a possible direction.

 Triple product amplitudes we derived \bref{sschrialamp}, \bref{twstedamp} is not the only way to construct 4pt graviton amplitude. We could try to construct it using a more general vertex operator. 
$$
V^{(n, m)}(z; k) = \epsilon^{(s)}_{\mu_{1}\cdots \mu_{n}}{}^{\nu_{1}\cdots \nu_{m}} \partial X_{L}^{\mu_{1}} \cdots X_{L}^{\mu_{n}}\bar{\partial} X_{R\;\nu_{1}} \cdots \bar{\partial} X_{R\;\nu_{m}} e^{ik\cdot (X_{R}(z) + X_{L}(\bar{z}))}
$$

$V^{(n, m)}(z; k)$ represents a $SO(d)\times SO(d)$ covariant state that is a product of spin-$n$ left-moving tensor and a spin-$m$ right moving tensor. 
After reduction to $SO(1, d-1)$ many representations of the Lorentz group will appear. If a spin-two symmetric tensor appears, then it could be interpreted as a graviton. For example, going back to original vertex operator with $n= m = 1$, after reduce this state to Lorentzian representation, we get $g_{\mu\nu}$ ( symmetric tensor), $B_{\mu\nu}$ (antisymmetric tensor) and $\phi$ (Dilaton). Vertex operator $V^{(1, 1)}$ corresponds to the NS-NS sector, not just gravitons.

For $V^{(n, m)}$ to contain gravitons, it needs to correspond to a massless state. This gives extra conditions on possible alternative vertex operators.  To systematically find the masses and exhaust all consistent vertex operators in a theory, one needs to develop a BRST prescription and find all BRST invariant vertex operators, which we will leave to future work.

If we try to calculate amplitudes using this vertex operator, one could get extra factors of Mandelstam variables from contractions between $\pd X^{\mu}$ and $e^{i k \cdot X}$ or contractions between $\pd X^{\mu}$ and $\pd X^{\nu}$, which will result in some polynomials multiplying original amplitude. Just as our deformation suggests.

\section{Higher-point amplitude}\label{sec:higher-point-amplitude}
In this section, we are going to more details on how to compute higher-point chiral string amplitudes, and eventually give an explicit calculation of 5-point amplitude. 

To get full amplitude, one needs to sum up all contributions of different topological configurations of defects. For $N$-point amplitude, there are $(N-1)!/2$ different configurations, which is just all distinct cyclic order of $N$-vertex operators up to a reflection. See figure \ref{pics1}.  

Amplitude for each configuration is calculated as follows: We started with the standard derivation of the KLT formula where we analytically continue complex-valued worldsheet coordinate $z$ to real-valued left and right-moving coordinate $z_{\pm}$. Thus, integration over worldsheet becomes integration over real lines in $z_{+}$ and $z_{-}$. See equation \bref{ampinte}.

 Integration over all left-moving modes can be decomposed into sums of all different vertex operators' orderings.\footnote{We first fixed ordering of all vertex operators \emph{then} use residual gauge symmetry to fix first, $N-1$th and $N$th vertex operator's position to $0, 1, \infty$.}
$$
\int_{\mathbb{R}} \prod_{i}^{N}dz_{i+}\bigr[\cdots\bigl] = \sum_{\sigma \in S_{N}}^{N}\int_{z_{\sigma(1)+}\leq \cdots z_{\sigma(N)+}}\prod_{i}^{N}dz_{i+}\bigr[\cdots\bigl].
$$
Each permutation of $\{z_{+i}\}$ corresponds to a contribution from different defect configurations.

\begin{figure}[h]
    \centering
    \import{.}{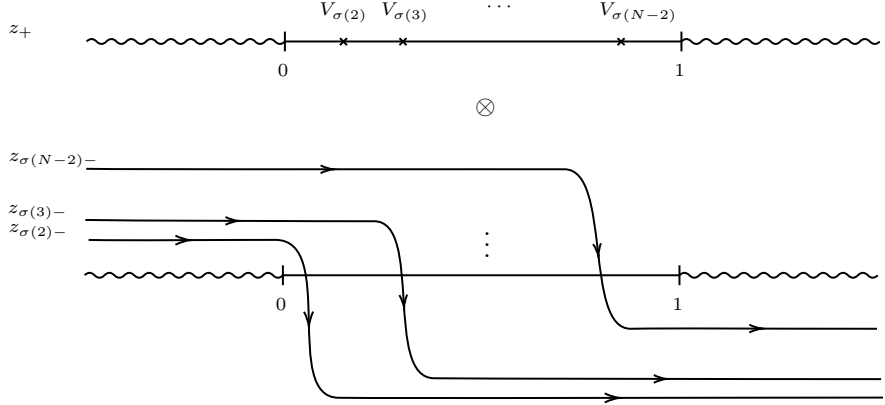}
        \caption{ A visual representation of integration contours in $z_{-}$, when positions of vertex operators in $z_{-}$ in the order $\{\sigma(1)\cdots \sigma(N)\}$ with $\sigma \in  S_{N}$.   \label{pics7}}
\end{figure}

Now, we need to perform integration over $z_{i-}$. Phase factor $\Phi(z_{1+}, z_{1-} \cdots z_{n-},  z_{n+})$ could be incorporated into integral naturally by specifying specific contours for each $z_{i-}$. See picture \ref{pics7}.

\FloatBarrier
As usual, we could ``fold'' these contours around cuts $(-\infty, 0]$ or $[1, \infty)$ which gives us sine factors in the KLT kernel. This step is essentially the same as the usual derivation of the KLT formula, so we won't elaborate on details here.  

After summing over all the different configurations, the final results are
\begin{equation}\label{mulklt}
M_{N} = \sum_{\sigma, \sigma{}'\in S_{N}/\mathbb{Z}_{N}} A_{N}^{(L)}(\sigma) S^{(R)}[\sigma|\sigma{}']A_{N}^{(R)}(\sigma{}'). 
\end{equation}
Here $S^{(R)}[\sigma|\sigma{}']$ is KLT kernel \cite{Bjerrum-Bohr:2010pnr}, but it depends on right-moving momentum. Hence, superscript $(\textrm{R})$ in its name. Similarly, superscript $(\textrm{L})$ and $(\textrm{R})$ in open string amplitude $A_N$ denote that whether amplitudes depend on left or right moving momenta. 

Without section condition, where left and right moving momenta are completely independent, this is the furthest one could simplify the result, but if section conditions are imposed, then further simplification occurs.

From monodromy relations, one could write the set of $(N-1)!$ open string amplitudes down in linear combination $(N-3)!$ independent ones. Without loss of generality, we choose the independent basis to be the set that fixes legs $1, N-1, N$. i.e.
$$
A_{N}(\beta(1, \cdots, N)\,) = \sum_{\sigma \in S_{N-3}}V[\beta|\sigma]A_{N}(1, \sigma(2, \cdots, N-2), N-1, N), \quad \beta\in S_N.
$$
$V[\beta|\sigma]$ is a $N! \times (N-3)!$ matrix where its entries are  rational function of $\sin(\pi \sum_{I}s_I)$. We use standard notation $(k_{i_1} + \cdots + k_{i_k})^2 = s_{i_1,\cdots i_k} \equiv s_I$ for kinematic invariants. 
 
Therefore, total amplitude could be written as (with at most $(N-3)!^2$ terms) 
$$
M_{N} = \sum_{\sigma, \sigma{}'\in S_{N-3}} A^{(L)}_{N}(1, \sigma(2, \cdots, N-2), N-1, N)\, \overline{S}[\sigma|\sigma{}']\, A^{(R)}_{N}(1, \sigma{}'(2, \cdots, N-2), N-1, N), 
$$
where 
$$\overline{S}[\sigma|\sigma{}'] =\sum_{\beta, \beta{}'\in S_{N}/\mathbb{Z}_N} V^{(L)}[\sigma| \beta]{}^{\top}\cdot S^{(R)}[\beta|\beta{}']\cdot V^{(R)}[\beta{}'| \sigma{}']{}$$
 In general, new kernel $\overline{S}[\beta|\beta{}']$ depends on both left and right-moving momenta, which makes its expression very complex,  and we would not necessarily gain any simplification.

But suppose that standard section condition \bref{sectioncond} is imposed, we have 
$s^{(L)}_{i_1,\cdots i_k} =  s^{(R)}_{i_1,\cdots i_k} +  2n$, $n$ is some integer. Therefore, we have $V^{(R)} = V^{(L)}$. Thus, $\overline{S}[\beta|\beta{}']$ only depends on the right-moving momenta and it is identical to the standard KLT kernel. 

$V^{(R)} = V^{(L)}$ also hold for twisted section condition \bref{sectioncond2}. To prove this, we need to show $V^{(R)}$ is invariant under parity transformation $s^{(R)}_{i_1, \cdots i_k} \rightarrow -s^{(R)}_{i_1, \cdots i_k}$. 
In order to see this, one needs to more explicit way of calculating $V^{(R)}$, which we will return to later.

Under either section conditions, the KLT formula for chiral strings could be written with at most $(N-3)!^2$ terms.
\begin{equation}\label{modNKLT}
M_{N} = \sum_{\sigma, \sigma{}'\in S_{N-3}} A^{(L)}_{N}(1, \sigma(2, \cdots), N-1, N)\, m^{-1\;(R)}[\sigma|\sigma{}']\, A^{(R)}_{N}(1, \sigma{}'(2, \cdots), N-1, N).
\end{equation}
This is the reason why 4-point amplitudes collapse to a single term when the section condition is imposed.

In formula above, we used inverse KLT kernel $m^{-1}[\beta|\beta{}'] \equiv S[\beta|\beta{}']$. This is because it is a more natural (and easy) object to calculate. Inverse KLT kernel could be calculated systematically using diagrammical rules similar to Feynman rules. \cite{Mizera:2016jhj} goes into detail on how it is done.

As mentioned earlier, we would like to show that $V[\beta|\sigma]$ is an even function of Mandelstam variables.
In the paper by Mizera \cite{Mizera:2016jhj}, they have shown that $V[\beta|\sigma]$ is related Inverse Kernel $m[\beta|\beta{}']$   
$$
V[\beta|\sigma]= m[\beta| \beta{}'] \cdot m^{-1}[\beta{}'|\sigma].
$$
The expression above implicitly sum over $\beta{}'$, which are  $(N-3)!$ permutations that fix $1, N-1, N$. i.e. $m[\beta| \beta{}'] $  is a $(N-1)!\times (N-3)!$ matrix while $m[\beta{}'|\sigma]$ is a $(N-3)!\times (N-3)!$ matrix.
 
If we follow the diagrammatical rules prescribed in \cite{Mizera:2016jhj} carefully, we can see that if $N$ is an even number, $m[\sigma|\sigma{}']$ is odd in Mandelstam variables, and if $N$ is odd, it is even in Mandelstam variables. Therefore, $V[\beta|\sigma]$ will always be an even function.

There is an easier way to see that $m[\sigma|\sigma{}']$ has parity $(-1)^{(N-1)}$. In the field theory limit where all Mandelstam variables much smaller than 1. The inverse kernel becomes a rational function, but the parity of $m[\sigma|\sigma{}']$  stays the same. In this case, the inverse kernel has a physical meaning, it is color-ordered amplitudes of $U(N)\times U(\tilde{N})$ bi-adjoint scalars \cite{Cachazo:2013gna, Cachazo:2013hca, Cachazo:2013iea}. We could write it schematically

$$
\tilde{m}[\sigma|\sigma{}'] = \pm \sum_{g\in {\cal G}_{\sigma}\cap{\cal G}_{\sigma{}'}} \frac{1}{\prod_{e\in g}s_{e}},
$$
where ${\cal G}_{\sigma}$ and ${\cal G}_{\sigma{}'}$ are the set of trivalent trees which is planar concerning ordering $\sigma$ and $\sigma{}'$ respectively. $s_{e}$ are Mandelstam variables that correspond to the square of momenta that flows through the internal edge $e$ of a trivalent tree $g$.

Since any trivalent tree with $N$ legs has $N-3$ internal edges, from the equation above we immediately shows that the parity of $\tilde{m}[\sigma|\sigma{}']$ (and thus  $m[\sigma|\sigma{}']$), is $(-1)^{N-1}$. 

\subsection{5-point amplitudes}\label{sec:5-point-amplitudes}
In this section, we will explicitly construct the 5-point chiral string amplitude using the modified KLT relation \bref{modNKLT}.

There are two independent 5-point open string amplitudes. We shall choose the basis $\{\,A_{\textrm{open}}(1, 2, 3, 4, 5), A_{\textrm{open}}(1, 3, 2,  4, 5)\,\}$. 

The $N$-point superstring disk amplitude could be written as a linear combination of $(N-3)!$ independent \emph{Yang-Mills} amplitudes multiplied by Gaussian hypergeometric functions \cite{Mafra:2011nv, Mafra:2011nw}.

\[
A_{\textrm{open}}(1,\cdots,  N) = \sum_{\sigma \in S_{N}}F(\sigma(2, \cdots, N-2) ) A_{\textrm{YM}}(1, \sigma(2, \cdots, N), N-1, N).
\]
$F(2, \cdots, N-2)$ is the Gaussian hypergeometric function which contributes massive poles to string amplitudes. It can be expressed as an integral. 
\[
F( 2, \cdots, n{-}2) = (-1)^{n-3} \int_{z_i < z_{i+1}} \prod^{n-2}_{j=2} dz_j \left( \prod |z_{i l}|^{s_{il}} \right)
\left(  \prod^{[n/2]}_{k=2} \sum^{k-1}_{m=1} {s_{mk}  \over z_{mk}} \right)
 \left( \prod^{n-2}_{k=[n/2]+1} \sum^{n-1}_{m=k+1} {s_{km}  \over z_{km}}   \right).
\]

Particularly for the 5-point case, we could write $F$ as a transformation matrix between two bases.
\[
\mathbf{A}_{\textrm{open}} = \mathbf{F}\cdot \mathbf{A}_{\textrm{YM}}, \quad 
\mathbf{F} \equiv 
\begin{bmatrix}
    F_1 & F_{2}\\
    F_3 & F_{4}
\end{bmatrix},
\]
where $F_{1} = F(2, 3)$, $F_{2} = F(3, 2)$, and  $F_{3}, F_{4}$ are the same integrals as $F_{1}, F_{2}$ but with momentum $2, 3$ exchanged.
\newpage
For 5-point amplitudes, these Gaussian hypergeometric integrals have closed form in terms of Generalized hypergeometric functions ${}_3F_{2}$. 
\footnote{We use the identity \cite{Slater_2008} 
\begin{align*}
&\int_{0}^{1}\int_{0}^{1}\, x^{a_1}y^{a_2}(1-x)^{a_3}(1-y)^{a_4}(1-xy)^{a_5}\, dxdy\\
 = &\pGq{1+a_1, 1+a_2, 1+a_3, 1+a_4}{2+a_1+a_3, 2+a_2+a_4} \pFq{3}{2}{1+a_1, 1+a_2, -a_5}{2+a_1 + a_3, 2+a_2+a_4}{1}.
\end{align*}}

\begin{align*}
F_1 & = s_{12}s_{34}\int_{0}^{1}\int_{0}^{1}\, (1-x)^{s_{34} -1 }x^{s_{45}}(1-y)^{s_{23}  }y^{s_{12}-1}(1-xy)^{s_{24}}dxdy\\
=& \pGq[16]{s_1 + 1, s_2+1, s_3+1, s_4 + 1}{s_1 + s_2 +  1, s_3 + s_4 + 1}\;\pFq[16]{3}{2}{s_1, s_4+1, s_2+s_3-s_5}{s_1 + s_2 +  1, s_3 + s_4 + 1}{1}\\
&\\
F_2 & = s_{13}s_{24}\int_{0}^{1}\int_{0}^{1}\, (1-x)^{s_{34} }x^{s_{45}}(1-y)^{s_{23}  }y^{s_{12}}(1-xy)^{s_{24} -1}dxdy\\
=& (s_4-s_2-s_1)(s_5-s_2-s_3)\pGq[16]{s_1 + 1, s_2+1, s_3+1, s_4 + 1}{s_1 + s_2 +  2, s_3 + s_4 + 2}\\
&\pFq[16]{3}{2}{s_1 + 1, s_4+1, s_2+s_3-s_5 + 1}{s_1 + s_2 +  2, s_3 + s_4 + 2}{1}
\end{align*}

\begin{align*}
F_3 & = s_{13}s_{24}\int_{0}^{1}\int_{0}^{1}\, (1-x)^{s_{24} -1 }x^{s_{45}}(1-y)^{s_{23}  }y^{s_{13}-1}(1-xy)^{s_{34}}dxdy\\
=& s_1s_3\pGq[16]{s_4 + 1,s_2+1, s_4-s_1-s_2 + 1, s_5-s_3-s_2 + 1}{s_4-s_1+  2, s_5+s_4-s_3-s_2+ 2}\\
&\pFq[16]{3}{2}{-s_3+1, s_4+1, s_4-s_1-s_2 + 1}{s_4-s_1+  2, s_5+s_4-s_3-s_2+ 2}{1}\\
&\\
F_4 & = s_{12}s_{34}\int_{0}^{1}\int_{0}^{1}\, (1-x)^{s_{24} }x^{s_{45}}(1-y)^{s_{23}  }y^{s_{13}}(1-xy)^{s_{34} -1}dxdy\\
=& \pGq[16]{s_2 + 1,s_4 + 1,s_4-s_1-s_2  + 1,s_5-s_3-s_2 + 1}{s_4-s_1+  1, s_5+s_4-s_3-s_2+ 1}\\
&\pFq[16]{3}{2}{-s_3, s_4+1, s_4-s_1-s_2}{s_4-s_1+  1, s_5+s_4-s_3-s_2+1}{1}.
\end{align*}

We have used the five independent kinematic invariants for 5-point scattering. $s_{1}\equiv s_{12}, s_{2} \equiv s_{23}, \cdots,  s_{5} \equiv s_{51}$, and we also use the notational shorthand 
$$
\pGq{a_1,  \cdots,  a_p}{b_1 , \cdots,  b_q} = \frac{\prod_{i = 1}^{p}\Gamma(a_i)}{\prod_{i = 1}^{q}\Gamma(b_i)}.
$$
\newpage
KLT kernel in $\{(12345), \; (13245)\}$ basis is
$$
\mathbf{S}_{\textrm{string}}  \equiv  S[\sigma|\sigma{}'] =  
\begin{pmatrix}
    S[\,I\,|\,I\,]&S[\, I |\, (23)\, ]\\
    S[\, (23)\, |\, I\,]&S[\, (23)\, |\, (23)\, ]
\end{pmatrix} \equiv 
\begin{pmatrix}
    S_1&S_3\\
    S_3&S_2
\end{pmatrix} ,
$$
where
\begin{align*}
    S_1 =&  \sin\pi s_1\sin \pi s_3\frac{\sin\pi(s_4-s_1)\sin\pi(s_2-s_4-s_5)\sin\pi(s_5-s_3) - \sin\pi s_1\sin \pi s_2 \sin \pi s_3}{\sin\pi(s_1-s_3-s_4)\sin\pi(s_3-s_1-s_5)\sin\pi(s_2-s_4-s_5)}\\
    &\\
    S_2 = & \frac{\sin\pi (s_4 - s_1 -s _2)\sin \pi (s_5-s_3-s_2)}{\sin\pi(s_1-s_3-s_4)\sin\pi(s_3-s_1-s_5)\sin\pi(s_2-s_4-s_5)}\\
    & (\sin\pi(s_1+s_2)\sin\pi(s_2+s_3)\sin\pi(s_2-s_4-s_5) - \sin\pi s_2\sin \pi (s_4-s_1-s_2) \sin \pi (s_5-s_1-s_3))\\
    &\\
 S_3 =& -\frac{\sin\pi s_1 \sin \pi s_3 \sin \pi (s_4-s_1-s_2)\sin \pi (s_5-s_2-s_3)\sin \pi (s_4+ s_5)}{\sin\pi(s_1-s_3-s_4)\sin\pi(s_3-s_1-s_5)\sin\pi(s_2-s_4-s_5)}.
\end{align*}

Putting everything together, we can represent the 5-point chiral string amplitude in terms of the bilinear of chiral string amplitudes. (We assume the section condition holds.)
$$
M_{5} = \mathbf{A_{\textrm{YM}}{}^{(L)}\cdot \Sigma\cdot A_{\textrm{YM}}{}^{(R)}}, \quad \mathbf{\Sigma} = \mathbf{\left[F^{(L)}\right]{}^{T}\cdot S_{\textrm{string}}{}^{(R)}\cdot F^{(R)}}.
$$
As usual, we use superscript $L$ and $R$ to denote whether to object should depend on left-moving or right-moving momenta.

One important exercise is to verify that in the special case $s_{i}^{(\textrm{L})} = -s_{i}^{(\textrm{R})}$, five-point chiral string amplitude indeed become graviton amplitude. This means that 
$$
\mathbf{\Sigma} \longrightarrow \mathbf{S_{\textrm{SURGA} }}, 
$$
where $\mathbf{S_{\textrm{SURGA} }}$ is the supergravity counterpart of KLT kernel. We could obtain it just by replacing $\sin \pi x $ with $x$ in $\mathbf{S}_{\textrm{string}}$, so it is a rational function of kinematic invariants.

Naively, $\mathbf{\Sigma}$ contains complicated products of ${}_3F_2$ and gamma functions, so it's very hard to imagine it would reduce to a rational function. But direct numerical verification  (By substituting kinematic invariants with random numbers) confirms that this is indeed the case.

Unfortunately, a direct proof that $ \mathbf{\Sigma} = \mathbf{S_{\textrm{SURGA} }} $ when $s_{i}^{(\textrm{L})} = -s_{i}^{(\textrm{R})}$ using properties of ${}_3F_2$ alone is beyond us. Proving this requires some non-trivial identities between products of generalized hypergeometric function at unit argument \cite{Slater_2008}.

\section{Discussion}
In this paper, find out that by simply making left and right-moving momenta independent of a bosonic worldsheet CFT, we could get a diverse set of new amplitudes. We call these kinds of worldsheet CFTs chiral string theories. 

Vertex operators in chiral string theories have non-trivial monodromies, we interpret this as vertex operators are now attached to defects. See figure \ref{pics1}. To get the amplitude, not only do we have to integrate over the moduli space, but we also have to sum over all the inequivalent topological configurations.

In particular, we find that if these left and right-moving amplitude satisfy section conditions \bref{sectioncond}, \bref{sectioncond2}, then chiral string amplitude could be represented in terms of a modified KLT formula \bref{modNKLT}. Where it is written as products of left-moving and right-moving open string amplitudes.

We also study 4-point chiral string amplitudes in detail. Assuming external states are represented as vertex operators of the kind \bref{gravexop}, we found that 4-point amplitudes admitted a triple product form. This kind of amplitude was initially proposed in \cite{Huang:2022mdb}. If section condition \bref{sectioncond} is imposed, then the triple product amplitude behaves like a string amplitude (indeed, the string amplitude is a special case of this), which contains an infinite number of resonance poles, but only finite of them exhibit infinite spin towers. 

On the other hand, when the twisted section condition is imposed, the triple product factor $F(s)$ becomes a rational function. There are three distinct cases.
(1) The amplitude is a pure polynomial with no poles. It represents theories with only higher derivative contact terms. (2) Amplitudes with a finite number of resonance poles, but each of them is an infinite spin tower. (3) The graviton amplitude.

The unitarity, UV-completeness, and Regge bound of these amplitudes are studied in detail. The only new amplitude that appears to be consistent with these conditions is \bref{sptwamp}. Although, as we mentioned, we assumed the special form of the vertex operator used, and a comprehensive search of vertex operators might reveal new consistent amplitudes.

We also go into discussions on how to calculate general $N$-point chiral string amplitude, where we found that if the section condition holds, the KLT kernel for the modified KLT formula is essentially the same as the KLT kernel for the usual ones, and we can represent the amplitude in the bilinear product of a $(N-3)! \times (N-3)!$ matrix. We also explicitly calculate the 5-point amplitude.

There are a few directly we would like to pursue from here.
\begin{enumerate}
    \item The explicit BRST quantization. If we could directly construct a BRST charge $Q_{\textrm{BRST}}$ from the worldsheet. It would tell what all the vertex operators in the theory should be, and what their masses are. (Imposing $Q_{\textrm{BRST}}V_{k} = 0$ would give constraint on $k^2$.) One might hope that we could use this formalism to prove unitarity rigorously similar to the Goddard-Thorn no-ghost theorem \cite{Goddard:1972iy}.
    \item Loop amplitude for chiral strings. The one-loop KLT relation was found in \cite{Stieberger:2022lss}. In that KLT formula, the left and right-moving momenta are already written separately. The generalization of this to a chiral string might not be too complicated. The modular invariance should put some restrictions on left and right-moving momenta. We should also check if one could reproduce loop graviton amplitudes using worldsheet theories, and how the UV divergence arises in this context.
    \item Chiral strings generalization of heterotic strings or other models. In heterotic strings, the left-moving modes are fundamentally different from the right-moving modes. So unlike the bosonic and type II strings, swapping the left and right moving momenta will result in fundamentally different theories.   
    \item Chiral strings without section condition or different section condition. It is also interesting to find the consistent set of amplitude without section condition. Most if not all of them contain infinite spin towers.
    \item  The unitarity constraints from the factorization of higer-point amplitudes gives additional constraints on the theory \cite{Arkani-Hamed:2023jwn}. Chiral string in general have non-polynomial multiparticle residues. For example $R_{n, m} \equiv \textrm{Res}{}_{s_{12} = n}\textrm{Res}{}_{s_{45} = m}\; M_5$ is not a polynomial of $s_{23}, s_{34}, s_{51}$ in general. This is just a manifestation of an infinite spin exchanges that we see in four point amplitude, but it makes tackling the consistent factorization much harder.
\end{enumerate}
\section*{Acknowledgment}
The author would like to thank Prof.\@ Yu-tin Huang and Prof.\@ Warren Siegel for the fruitful discussions on this subject.


\printbibliography

\end{document}